\documentclass[aps,preprint,prd,showpacs,nofootinbib]{revtex4-1}
\pdfoutput=1

\usepackage[utf8]{inputenc}
\usepackage{amsmath}
\usepackage{bm}
\usepackage{graphicx}
\usepackage{xcolor}
\usepackage{float}
\usepackage{longtable}
\usepackage{multirow}
\usepackage{array}
\usepackage{tabularx}
\usepackage{diagbox}
\usepackage{scalerel}

\usepackage{appendix}
\usepackage{ulem}
\usepackage{soul}
\usepackage{footmisc}
\usepackage{wasysym}
\usepackage{lineno}
\usepackage{fancyhdr}

\usepackage[colorlinks=true,linkcolor=blue,citecolor=blue]{hyperref}

\newcommand{\be}{\begin{equation}}
\newcommand{\ee}{\end{equation}}
\newcommand{\bea}{\begin{eqnarray}}
\newcommand{\eea}{\end{eqnarray}}
\newcommand{\bes}{\begin{subequations}}
\newcommand{\ees}{\end{subequations}}

\newcommand{\bc}{\begin{center}}
\newcommand{\ec}{\end{center}}

\begin{document}
\title{Three decades of FCNC studies in 3-3-1 model with right-handed neutrinos: from \texorpdfstring{$Z^\prime$}{Z'} - dominance to the alignment limit}
\author{Patricio Escalona$^1$}\email{patricioescalona96@gmail.com}
\author{João Paulo Pinheiro$^{2\,,3}$}
\author{Vinícius Oliveira$^{4\,,5\,,6}$}
\author{A. Doff$^7$}
\author{C. A. de S. Pires$^1$}
\affiliation{$^1$Departamento de F\'isica, Universidade Federal da Para\'iba, Caixa Postal 5008, 58051-970, Jo\~ao Pessoa, PB, Brazil} 
\affiliation{$^2$State Key Laboratory of Dark Matter Physics, Tsung-Dao Lee Institute \& School of Physics and Astronomy, Shanghai Jiao Tong University, Shanghai 200240, China}
\affiliation{$^3$Key Laboratory for Particle Astrophysics and Cosmology (MOE) \& Shanghai Key Laboratory for Particle Physics and Cosmology, Shanghai Jiao Tong University, Shanghai 200240, China}
\affiliation{$^4$Departamento de Física da Universidade de Aveiro and \\
  Center for Research and Development in Mathematics and Applications (CIDMA),\\
  Campus de Santiago, 3810-183 Aveiro, Portugal}
\affiliation{$^5$Laboratório de Instrumentação e Física Experimental de Partículas (LIP), \\
  Universidade do Minho, 4710-057 Braga, Portugal}
\affiliation{$^6$Department of Physics, Lund University, 221 00 Lund, Sweden}
\affiliation{$^7$Universidade Tecnologica Federal do Parana - UTFPR - DAFIS, R. Doutor Washington Subtil Chueire, 330 - Jardim Carvalho,  84017-220, Ponta Grossa, PR, Brazil} 

\begin{abstract} 
Flavor-changing neutral current (FCNC) processes play a prominent role in the search for physics beyond the Standard Model (SM) due to their sensitivity to new physics at the TeV scale. Meson-antimeson transitions and rare meson decays provide stringent constraints on new physics through precision measurements of observables such as mass differences, CP asymmetries, and branching ratios. Extensions of the SM based on the $\text{SU}(3)_C \times \text{SU}(3)_L \times \text{U}(1)_N$ gauge group offer a compelling framework for flavor physics, as FCNC processes emerge inexorably at tree level due to the non-universal transformations of the quark families. Among its various realizations, the version incorporating right-handed neutrinos (331RHN) is the most phenomenologically viable. This review synthesizes three decades of theoretical developments in FCNC phenomenology within the 331RHN model, from early $Z^\prime$-dominated studies to the recent recognition of the decisive role played by the SM-like Higgs boson and the identification of the alignment limit. We demonstrate that viable parameter space spans orders of magnitude---from $m_{Z^\prime} \sim$ a few hundred GeV to $\sim 100$ TeV---depending critically on quark mixing parametrizations and scalar alignment configurations, with significant implications for experimental searches at current and future colliders.
\end{abstract}
\maketitle

\tableofcontents

\section{Introduction}

Flavor-changing neutral current (FCNC) processes play a central role in the search for physics beyond the Standard Model (SM)~\cite{Altmannshofer:2025rxc}. Meson--antimeson transitions such as $K^0 -\bar K^0 \,,\, D^0- \bar D^0\,,\, B^0_d -\bar B^0_d$ and $B^0_s -\bar B^0_s$~\cite{Nierste:2025jxe}, along with rare meson decays, provide stringent probes of new physics at the TeV scale\footnote{In this review, we will usually omit the superscript $0$ because the focus is mainly on neutral mesons.}. These processes are being intensively studied at the LHCb~\cite{LHCb:2017qna,LHCb:2021dcr,LHCb:2021awg,LHCb:2022qnv, LHCb:2022vje} and Belle II~\cite{Belle-II:2018jsg,Browder:2021hbl} experiments, building upon pioneering work at CLEO~\cite{CLEO:1991qyy}, Belle~\cite{Belle:2017oht,Belle:2000cnh}, and BABAR~\cite{BaBar:1998yfb,BaBar:2001yhh}. The sensitivity of FCNC observables to beyond-the-SM (BSM) physics arises from a fundamental feature of the SM: the Glashow--Iliopoulos--Maiani (GIM) mechanism~\cite{Glashow:1970gm} forbids FCNC at tree level, relegating them to highly suppressed loop-induced effects through penguin and box diagrams. Any tree-level contribution from new physics can therefore leave detectable imprints, even if the associated energy scale lies well beyond direct collider reach.

Among BSM frameworks, models based on the $\text{SU}(3)_C \times \text{SU}(3)_L \times \text{U}(1)_N$ gauge group---collectively known as 3-3-1 models~\cite{Schechter:1973nqg,Schechter:1974mq,Clavelli:1974ez,Gupta:1973pv,Segre:1976rc,Fritzsch:1976dq,Kandaswamy:1977ct,Lee:1977qs,Buccella:1977gx,Lee:1977tx,Langacker:1977ae,Sutherland:1978vw,Komatsu:1978jy,Moriya:1977cx,Fischer:1978zu,Dahmen:1978fv,Singer:1978zd,Singer:1980sw,Pisano:1992bxx,Frampton:1992wt,Montero:1992jk,Foot:1994ym,Ozer:1995xi}---offer a particularly compelling structure for flavor physics. In these models, anomaly cancellation combined with asymptotic freedom naturally explains the existence of exactly three fermion families.  Crucially, anomaly cancellation requires that one quark family transform as a triplet under $\text{SU}(3)_L$ while the other two transform as anti-triplets~\cite{Frampton:1992wt,Pisano:1996ht,Liu:1993gy}. This non-universal quark assignment \textit{inevitably} generates tree-level FCNC mediated by both the $Z^\prime$ gauge boson and neutral scalars, including the SM-like Higgs boson. Among the various realizations of 3-3-1 models, we focus on the version incorporating right-handed neutrinos (331RHN)~\cite{Montero:1992jk,Foot:1994ym}, which avoids the Landau pole problem that afflicts the minimal 3-3-1 model below the symmetry-breaking scale~\cite{Dias:2004dc,Dias:2004wk,Doff:2023bgy}.

The phenomenology of FCNC in 331RHN models has evolved considerably over three decades of investigation. Early studies in the late 1990s and 2000s focused predominantly on $Z^\prime$-mediated contributions~\cite{Long1:1999lv,Long2:2001lv,Sher:2004Sr,Ochoa:2008Mo}, typically adopting quark mixing  parametrizations inspired by, e.g., the Fritzsch ansatz. These analyses obtained lower bounds on the $Z^\prime$ mass ranging from a few TeV to tens of TeV, depending on the assumed flavor structure. However, a critical limitation of this early work was the systematic \textit{neglect} of scalar contributions to FCNC---an oversimplification that, as recent work has shown, can lead to misleading conclusions about the viable parameter space.

The recognition that neutral scalars play an important role began in the 2010s~\cite{Farinaldo1:2012Df,Okada:2016whh}, but most studies continued to treat scalar and gauge contributions separately. A major conceptual advance came with the realization that the SM-like Higgs boson, being the lightest neutral scalar at $m_h = 125$ GeV, generically provides the \textit{dominant} contribution to meson transitions~\cite{Oliveira:2022dav}. Unlike $Z^\prime$ contributions, which can be suppressed by increasing the gauge boson mass, scalar-mediated FCNCs scale as $1/m_h^2$ and are tied to quark masses and Cabibbo--Kobayashi--Maskawa parameters. This poses a fundamental challenge: any proposed parametrization of the quark mixing matrices must be tested against the SM-like Higgs contribution to ensure consistency with experimental limits on $\Delta M_M$, the mass difference between a meson $M=(D,K,B_d,B_s)$ and its antiparticle, which characterizes the transition frequency.

The most recent and impactful development is the identification of the \textit{alignment limit}~\cite{Escalona:2025rxu}---a specific configuration of scalar mixing angles that naturally suppresses FCNC mediated by the SM-like Higgs. Within this limit, two different scenarios emerge depending on whether FCNC arises in the up-quark sector (allowing $Z^\prime$ masses as low as a few hundred GeV) or the down-quark sector (requiring $Z^\prime$ masses in the hundreds of TeV range). This discovery fundamentally changes our understanding of the 331RHN parameter space and clarifies the relationship between scalar alignment, quark mixing patterns, and experimental constraints from neutral meson systems.

This review provides a comprehensive overview of FCNC phenomenology in the 331RHN model, synthesizing three decades of theoretical progress. We emphasize throughout that any meaningful analysis of FCNC in 3-3-1 models must account for contributions from all neutral mediators---gauge bosons and scalars alike---with particular attention to the role of the SM-like Higgs. The review is organized as follows. Section~\ref{sec:essence} introduces the 331RHN model, detailing the scalar and gauge sectors. Section~\ref{sec:structure} discusses the general structure of tree-level FCNC and explains how different choices of which quark family transforms as a triplet lead to distinct model variants. Section~\ref{sec:variants} presents explicit FCNC interactions and meson transition formulae for each variant. Section~\ref{sec:state_of_art} traces the historical development of FCNC studies in 331RHN, from early $Z^\prime$-dominated analyses through the gradual incorporation of scalar effects up to the recent alignment limit framework, while also covering family discrimination, $Z$-$Z^\prime$ mixing constraints, and updated bounds on new physics masses.  Finally, Section~\ref{sec:conclusion} synthesizes the main lessons learned over three decades, discusses open theoretical questions including the physical origin of the alignment limit and the role of right-handed quark mixing, and outlines experimental prospects across different mass scales---from direct $Z^\prime$ searches at the HL-LHC to ultra-precision flavor measurements at Belle II and future kaon experiments. We conclude by emphasizing that the interplay between collider searches and precision flavor physics will be decisive in testing the 331RHN framework in the coming decade.
\section{The essence of the 331RHN}\label{sec:essence}

The 331RHN addresses fundamental questions left unanswered by the SM, such as the origin of fermion families, electric charge quantization, and neutrino masses. In this section, we review the essential features of the 331RHN model.

In $\text{SU}(3)_C \times \text{SU}(3)_L \times \text{U}(1)_N$  models, the electric charge operator takes the form
\begin{equation}
    \frac{\hat Q}{e}=\frac{1}{2}(\lambda_3 +\beta \lambda_8) +NI,
\end{equation}
where $e$ is the electron electric charge, $\lambda_3$ and $\lambda_8$ are the diagonal Gell-Mann matrices,  $\beta$ is an embedding parameter, $N$ is a free quantum number and $I$ is the identity operator. The value of $\beta$ determines the specific version of the 3-3-1 model. For example,  $\beta=-\sqrt{3}$ corresponds to  the minimal 3-3-1 model, whereas the 331RHN requires  $\beta=-\frac{1}{\sqrt{3}}$. 

Once $\beta$ is fixed, the electric charge operator  $\hat Q$ becomes  a function of the quantum number $N$. For the 331RHN model, the relation between the electric charge operator and the group generators is: $$ \frac{\hat Q}{e}=\frac{1}{2}(\lambda_3 -\frac{1}{\sqrt{3}} \lambda_8) +NI,$$  which implies the following electric charges for the fermionic multiplets:
\begin{eqnarray}
   && \mbox{Triplets}\to \left (
\begin{array}{c}
 \frac{1}{3}+N\\
 -\frac{2}{3} +N\\
\frac{1}{3}+N
\end{array}
\right )e,\nonumber \\
&&  \mbox{Anti-triplets}\to \left (
\begin{array}{c}
 -\frac{1}{3}+N\\
 \frac{2}{3} +N\\
-\frac{1}{3}+N
\end{array}
\right )e, \nonumber \\
&&  \mbox{Singlets} \to Ne .
\end{eqnarray}
Thus, by fixing the values of $N$, we automatically determine the electric charge pattern of the model’s particles. 

It was  shown in Ref.~\cite{deSousaPires:1998jc,deSousaPires:1999ca} that, in any 3-3-1 model, the invariance of Yukawa interactions, combined with anomaly cancellation, provides the constraints needed to fix all $N$ values. Consequently, 3-3-1 models  offer an explanation for the pattern of  electric charge quantization. For example, for the lepton triplet, the approach discussed above provides $N_{l_L}=-\frac{1}{3}$, while for the lepton singlet, $N_{l_R}=-1$, which implies the following electric charges for the leptons:
\begin{eqnarray}
   && \mbox{Lepton triplet}\to \left (
\begin{array}{c}
 0\\
 -1 \\
0
\end{array}
\right )e \,\,\,\,\,\,\ \to \,\,\,\,\, L_{l} =\left (
\begin{array}{c}
\nu_{l_L} \\
e_{l_L} \\
(\nu_{l_R})^{c}
\end{array}
\right ) \sim(1\,,\,3\,,\,-1/3), \nonumber \\
&&  \mbox{Lepton singlet }\to -1e\,\,\,\,\,\,\,\,\,\,\, \to \,\,\,\,\,e_{l_R}\sim(1,1,-1),
\end{eqnarray}
with $l=e,\mu,\tau$ labeling the three lepton families.

Following the same approach for the quark sector (which we will discuss in more detail in Section ~\ref{sec:structure} and Section ~\ref{sec:variants}), the quantum number $N$ acquires the value $N_Q=\frac{1}{3}$ for the quark triplet, giving the following charge distribution:
\begin{equation}
\mbox{Quark triplet}\to \left ( \begin{array}{c}
+\frac{2}{3} \\
-\frac{1}{3} \\
+\frac{2}{3}
\end{array}
\right )e,
\end{equation}
while for the anti-triplet of quarks, we have $N_{\bar Q}=0$, giving the following electric charges:
\begin{equation}
 \mbox{Quark anti-triplet}\to\left ( \begin{array}{c}
-\frac{1}{3} \\
+\frac{2}{3} \\
-\frac{1}{3}
\end{array}
\right )e.    
\end{equation}

For the singlet of quarks, invariance of the Yukawa terms and anomaly cancellation provide $N_{q_R}=\frac{1}{3}$ and $N_{q_R}=\frac{2}{3}$. Observe that, in the 331RHN, the new quarks have standard electric charges. The same approaches applies to the scalar and gauge sectors. 

We now turn to the scalar sector, crucial for symmetry breaking and mass generation. The model contains three scalar triplets~\cite{Montero:1992jk,Foot:1994ym},
\begin{eqnarray}
\eta = \left (
\begin{array}{c}
\eta^0 \\
\eta^- \\
\eta^{\prime 0}
\end{array}
\right ),\,\rho = \left (
\begin{array}{c}
\rho^+ \\
\rho^0 \\
\rho^{\prime +}
\end{array}
\right ),\,
\chi = \left (
\begin{array}{c}
\chi^0 \\
\chi^{-} \\
\chi^{\prime 0}
\end{array}
\right ),
\label{scalarcont} 
\end{eqnarray}
where $\eta$ and $\chi$ transform as $(1\,,\,3\,,\,-1/3)$ and $\rho$ as $(1\,,\,3\,,\,2/3)$. 

The most economical pattern of spontaneous symmetry breaking occurs when only $\eta^0$, $\rho^0$ and $\chi^{\prime 0}$ acquire nonzero vacuum expectation values (VEVs), namely $v_\eta$, $v_\rho$ and $v_{\chi^{\prime}}$, respectively. For simplicity, we will assume this configuration in what follows. Scenarios where all neutral scalars acquire VEVs are also interesting but more complex since the standard quarks mix with the new ones~\cite{Doff:2006rt}. 

The scalar spectrum is obtained from the potential
\begin{eqnarray} 
V(\eta,\rho,\chi)&=&\mu_\chi^2 |\chi|^2 +\mu_\eta^2|\eta|^2
+\mu_\rho^2|\rho|^2+\lambda_1|\chi|^4 +\lambda_2|\eta|^4
+\lambda_3|\rho|^4+ \nonumber \\
&&\lambda_4(\chi^{\dagger}\chi)(\eta^{\dagger}\eta)
+\lambda_5(\chi^{\dagger}\chi)(\rho^{\dagger}\rho)+\lambda_6
(\eta^{\dagger}\eta)(\rho^{\dagger}\rho)+ \nonumber \\
&&\lambda_7(\chi^{\dagger}\eta)(\eta^{\dagger}\chi)
+\lambda_8(\chi^{\dagger}\rho)(\rho^{\dagger}\chi)+\lambda_9
(\eta^{\dagger}\rho)(\rho^{\dagger}\eta) \nonumber \\
&&-\frac{f}{\sqrt{2}}\epsilon^{ijk}\eta_i \rho_j \chi_k +\mbox{H.c.}\,,
\label{potential}
\end{eqnarray}
shifting the neutral scalar fields as
\begin{equation} 
\eta^0 \,,\,\rho^0 \,,\,\chi^{\prime 0}  \to \frac{1}{\sqrt{2}}(v_{\eta\,,\,\rho\,,\,{\chi^{\prime}}} + R_{\eta\,,\,\rho\,,\,{\chi^{\prime}}} + iI_{\eta\,,\,\rho\,,\,{\chi^{\prime}}}).
\end{equation}
The above potential is the most economical one~\cite{Pal:1994ba} which conserves the lepton number. Moreover, it is invariant under a $Z_2$ discrete symmetry defined as~\footnote{This symmetry avoids Yukawa terms that lead to mixing between the standard and exotic quarks, considered in~\cite{Doff:2006rt}.}
\begin{equation}
\eta\,,\,\rho\,,\,e_{l_R} \,,\,u_{a_R} \,,\, d_{a_R}  \to -(\eta\,,\,\rho\,,\,e_{l_R} \,,\,u_{a_R} \,,\, d_{a_R} ).
\end{equation}
Substituting the above shifts into the potential, we obtain the conditions ensuring a global minimum of the potential\footnote{For studies of potential stability and the global minimum, see~\cite{Sanchez-Vega:2018qje,Dorsch:2024ddk,Kannike:2025qru}.}:
\begin{eqnarray}
 &&\mu^2_\chi +\lambda_1 v^2_{\chi^{\prime}} +
\frac{\lambda_4}{2}v^2_\eta  +
\frac{\lambda_5}{2}v^2_\rho-\frac{f}{2}\frac{v_\eta v_\rho}
{ v_{\chi^{\prime}}}=0,\nonumber \\
&&\mu^2_\eta +\lambda_2 v^2_\eta +
\frac{\lambda_4}{2} v^2_{\chi^{\prime}}
 +\frac{\lambda_6}{2}v^2_\rho -\frac{f}{2}\frac{v_{\chi^{\prime}} v_\rho}
{ v_\eta} =0,
\nonumber \\
&&
\mu^2_\rho +\lambda_3 v^2_\rho + \frac{\lambda_5}{2}
v^2_{\chi^{\prime}} +\frac{\lambda_6}{2}
v^2_\eta-\frac{f}{2}\frac{v_\eta v_{\chi^{\prime}}}{v_\rho} =0.
\label{mincond} 
\end{eqnarray}

To reproduce the SM electroweak scale, we will take $v_\eta^2 + v_\rho^2 = v_{\text{SM}}^2$, where $v_{\text{SM}} = 246~\text{GeV}$ (see Eq.~\ref{recoveringSMvev}), while $v_{\chi^{\prime}}$ sets the $\text{SU}(3)_L \times \text{U}(1)_N$ breaking scale. Moreover, current LHC constraints require $v_{\chi^{\prime}} \gtrsim 10~\text{TeV}$~\cite{Coutinho:2013lta,Alves:2022hcp}, implying that $v_{\chi^{\prime}} \gg v_\eta, v_\rho$. This hierarchy will play an important role in the discussion below.

The scalar mass matrices can be obtained only after solving the potential’s minimization condition. The mass matrix for the CP-even neutral scalars in the basis $(R_{\chi'}, R_\eta, R_\rho)$ is
\begin{equation}
M_R^2=
\begin{pmatrix}
\lambda_1 v^2_{\chi^{\prime}}+fv_\eta v_\rho/4v_{\chi^{\prime}} & \lambda_4 v_{\chi^{\prime}}  v_\eta/2- f v_\rho/4 &  \lambda_5 v_{\chi^{\prime}}  v_\rho/2- f v_\eta/4\\
\lambda_4 v_{\chi^{\prime}}  v_\eta/2- f v_\rho/4 & \lambda_2 v^2_\eta+ fv_{\chi^{\prime}} v_\rho/4v_\eta & \lambda_6 v_\eta v_\rho/2-f v_{\chi^{\prime}}/4\\
\lambda_5 v_{\chi^{\prime}}  v_\rho/2- f v_\eta/4 &  \lambda_6 v_\eta v_\rho/2-f v_{\chi^{\prime}}/4 &  \lambda_3 v^2_\rho+ fv_{\chi^{\prime}} v_\eta/4v_\rho
\end{pmatrix}.
\label{masseven}
\end{equation}
In what follows, we adopt the decoupling limit\footnote{This decoupling limit is physically motivated when $v_{\chi'} \gg v_\eta, v_\rho$,  ensuring that heavy 3-3-1 physics decouples from the electroweak scale.}, which amounts to decoupling $R_{\chi'}$ from $R_\eta$ and $R_\rho$. This requires
\begin{equation}
     \lambda_4 v_{\chi^{\prime}}  v_\eta/2- f v_\rho/4=0\, \quad \text{ and }\, \quad \lambda_5 v_{\chi^{\prime}}  v_\rho/2- f v_\eta/4=0.
\end{equation}
In this limit, the mass matrix reduces to
\begin{equation}
M_H^2=
\begin{pmatrix}
 \lambda_2 v^2_\eta+ fv_{\chi^{\prime}} v_\rho/4v_\eta & \lambda_6 v_\eta v_\rho/2-f v_{\chi^{\prime}}/4\\
  \lambda_6 v_\eta v_\rho/2-f v_{\chi^{\prime}}/4 &  \lambda_3 v^2_\rho+ fv_{\chi^{\prime}} v_\eta/4v_\rho
\end{pmatrix},
\label{masseven2}
\end{equation}
in the basis $( R_\eta,R_\rho)$. 
After diagonalization, the CP-even eigenstates are
\begin{eqnarray}
    &&h=\cos \varphi R_\eta +  \sin \varphi R_\rho, \nonumber \\
    && H= \cos \varphi R_\rho -  \sin \varphi R_\eta,\nonumber \\
   && H^{\prime}= R_{\chi^{\prime}}.
\end{eqnarray}
Here, $H'$ is a heavy CP-even scalar with mass set by $v_{\chi'}$. The remaining CP-even scalars are $h$ (the SM-like Higgs) and $H$, whose mass depends on $f$ and $v_{\chi'}$. The angle $\varphi$ diagonalizes $M_H^2$ via a $2\times2$ rotation. For the explicit diagonalization procedure, we refer the reader to Ref.~\cite{Dong:2006mg,CatoPires:2006hy}.

Next, we consider the CP-odd scalars. Considering the basis $(I_{\chi^{\prime}}, I_\eta, I_\rho)$, the mass matrix is 
\begin{equation}
M_I^2=\frac{1}{4}
\begin{pmatrix}
fv_\eta v_\rho/v_{\chi^{\prime}} &  f v_\rho &  f v_\eta\\
f v_\rho &  fv_{\chi^{\prime}} v_\rho v_\eta & f v_{\chi^{\prime}}\\
 f v_\eta &  f v_{\chi^{\prime}} &  fv_{\chi^{\prime}} v_\eta/v_\rho
\end{pmatrix}.
\label{MI}
\end{equation}
This matrix can be diagonalized analytically.  Assuming $v_{\chi^{\prime}}\gg v_\eta\,,\,v_\rho$, we obtain two zero eigenvalues, which correspond to the eigenstates $I_{\chi^{\prime}}$ and $G= \frac{v_\eta}{\sqrt{v^2_\eta + v^2_\rho}} I_\eta -\frac{v_\rho}{\sqrt{v^2_\eta + v^2_\rho}}I_\rho$, and a massive eigenstate $A= \frac{v_\rho}{\sqrt{v^2_\eta + v^2_\rho}} I_\eta +\frac{v_\eta}{\sqrt{v^2_\eta + v^2_\rho}}I_\rho$ with mass given by  $m^2 _A=\frac{f v_{\chi^{\prime}}}{4}(\frac{v_\eta v_\rho}{v^2_{\chi^{\prime}}}+\frac{v_\eta}{v_\rho}+\frac{v_\rho}{v_\eta})$. We define $\sin \phi= \frac{v_\rho}{\sqrt{v^2_\eta + v^2_\rho}}$ and $\cos \phi = \frac{v_\eta}{\sqrt{v^2_\eta + v^2_\rho}}$. Note that if $f=0$, then $m_A=0$, meaning that $A$ is a pseudo-Goldstone boson associated with some U$(1)_X$ global symmetry. In fact, it was shown in Ref.~\cite{Pal:1994ba} that U$(1)_X$ is the Peccei--Quinn symmetry and $A$ is the Weinberg--Wilczek axion~\cite{Weinberg:1977ma,Wilczek:1977pj}. This axion must, of course, be avoided. The trilinear term $f \eta \rho \chi$ minimally breaks the Peccei--Quinn symmetry while preserving lepton number\footnote{For the implementation of Peccei--Quinn symmetry with a viable axion in the 331RHN, see~\cite{Dias:2003iq}.}.

The 331RHN has two other neutral scalars, $\chi^0$ and $\eta^{\prime 0}$. Both carry two units of lepton number~\cite{deSPires:2007wat}. Assuming lepton number conservation, they do not mix with $\eta^0$, $\rho^0$, or $\chi^{\prime 0}$. In the  basis $(\chi^{ 0} , \eta^{\prime 0})$ the mass matrix reads:
\begin{equation}
M^2_{\chi \eta^{\prime}}=\frac{1}{4}
\begin{pmatrix}
\lambda_7v^2_\eta +f v_\eta v_\rho/v_{\chi^{\prime}} & -\lambda_7 v_\eta v_{\chi^{\prime}}-fv_\rho\\
-\lambda_7 v_\eta v_{\chi^{\prime}}-fv_\rho &  \lambda_7 v^2_{\chi^{\prime}} +f v_\rho v_{\chi^{\prime}}/v_\eta
\end{pmatrix}.
\label{MI2}
\end{equation}

After diagonalizing it, we obtain  $G_3=\frac{v_{\chi^{\prime}}}{\sqrt{v^2_\eta + v^2_{\chi^{\prime}}}}\chi^0 +\frac{v_\eta}{\sqrt{v^2_\eta + v^2_{\chi^{\prime}}}}\eta^{\prime 0}$, which is a Goldstone boson absorbed by the non-Hermitian gauge bosons $U^0$ and $U^{0 \dagger}$. The other neutral scalar is  $G_4=-\frac{v_\eta}{\sqrt{v^2_\eta + v^2_{\chi^{\prime}}}}\chi^0 +\frac{v_{\chi^{\prime}}}{\sqrt{v^2_\eta + v^2_{\chi^{\prime}}}}\eta^{\prime 0}$, with mass $m_{G_4}^2=\frac{\lambda_7}{4}(v_{\chi^{\prime}}^2+ v^2_\eta)+\frac{fv_\rho}{4}(\frac{v_{\chi^{\prime}}}{v_\eta}+\frac{v_\eta}{v_{\chi^{\prime}}})$. This mass  depends on $f$ and $v_{\chi^{\prime}}$, which means that $G_4$ is a heavy particle regardless of the value of $f$. Moreover, due to lepton number conservation, and because $G_4$ is a bilepton, its interactions must involve another bileptons of the model such as the new quarks, $W^{\prime \pm}$, $U^0$, or $\rho^{\prime \pm}$. If $G_4$ is the lightest bilepton, it is stable and can serve as a dark matter candidate~\cite{deSPires:2007wat,Mizukoshi:2010ky}.

We now discuss the charged scalars. In the basis $(\chi^- , \rho^{\prime -}, \eta^- , \rho ^-)$, the mass matrix is given by:
\begin{equation}
M_C^2=\frac{1}{2}
\begin{pmatrix}
\lambda_8 v^2_\rho+f v_\eta v_\rho / v_{\chi^{\prime}} & \lambda_8 v_\rho v_{\chi^{\prime}} +fv_\eta & 0 & 0 \\
\lambda_8 v_\rho v_{\chi^{\prime}} +fv_\eta & \lambda_8 v^2_{\chi^{\prime}} +f v_\eta v_{\chi^{\prime}}/ v_\rho & 0 & 0\\
0 & 0 & \lambda_9 v^2_\rho  +f v_\rho v_{\chi^{\prime}}/ v_\eta & \lambda_9 v_\rho v_\eta  +f  v_{\chi^{\prime}} \\
0 & 0 &  \lambda_9 v_\rho v_\eta  +f  v_{\chi^{\prime}} & \lambda_9 v^2_\eta  +f v_\eta v_{\chi^{\prime}}/ v_\rho
\end{pmatrix}.
\label{M+}
\end{equation}
We note that $\chi^-$ and $\rho^{\prime -}$ also carry two units of lepton number each~\cite{deSPires:2007wat}. Thus, lepton number conservation prevents their mixing with $\eta^- , \rho ^-$. After diagonalizing this matrix we obtain two Goldstone bosons: $G_1^+=-\frac{v_\rho}{\sqrt{v^2_\eta + v^2_\rho}}\rho^+ +\frac{v_\eta}{\sqrt{v^2_\eta + v^2_\rho}} \eta^+$ and $ G_2^{ +}= \frac{v_{\chi^{\prime}}}{\sqrt{v^2_\eta + v^2_{\chi^{\prime}}}}\chi^+ +\frac{v_{\eta}}{\sqrt{v^2_\eta + v^2_{\chi^{\prime}}}}\rho^{\prime +}$, which are absorbed by $W^{\pm}$ and $W^{\prime \pm}$. The remaining two heavy charged scalars are $h_1^+=\frac{v_\eta}{\sqrt{v^2_\eta + v^2_\rho}}\rho^+ +\frac{v_\rho}{\sqrt{v^2_\eta + v^2_\rho}} \eta^+ $ with mass $m_{h_1^{\pm}}^2=\frac{1}{2}(f v_{\chi^{\prime}}+\lambda_9 v_\eta v_\rho)(\frac{v_\eta}{v_\rho}+\frac{v_\rho}{v_\eta}) $  and $h_2^+=-\frac{v_{\eta}}{\sqrt{v^2_\eta + v^2_{\chi^{\prime}}}}\chi^++ \frac{v_{\chi^{\prime}}}{\sqrt{v^2_\eta + v^2_{\chi^{\prime}}}}\rho^{\prime +} $   with mass  $m^2_{h_2^+}=\frac{1}{2}(f v_{\eta}+\lambda_8 v_\rho v_{\chi^{\prime}})(\frac{v_{\chi^{\prime}}}{v_\rho} +\frac{v_\rho}{v_{\chi^{\prime}}}) $. The dependence of $m^2_{h_2^{\pm}}$ on $f$ and $v_{\chi^{\prime}}$ ensures that $h_2^{\pm}$ remains heavy, similarly to $G_4$ discussed above.

In summary, the scalar spectrum of the 331RHN model comprises nine physical states beyond the SM. Apart from the SM-like Higgs boson $h$, the scalar spectrum contains four heavy scalars ($H^\prime$, $G_4$, and $h^{\pm}_2$) with masses set by the 3-3-1-breaking scale $v_{\chi'}$, and four additional scalars ($H$, $A$, and $h^{\pm}_1$) whose masses are determined by the trilinear coupling $f$. The role of this coupling in explicitly breaking the Peccei--Quinn symmetry makes the phenomenology of the model strongly dependent on its value~\cite{Pinheiro:2022yex}.

Two distinct regimes emerge: (i) When $f \lesssim v_{\eta}, v_{\rho}$, the scalars $h$, $H$, $A$, and $h^{\pm}_1$ form a light spectrum at the electroweak scale, effectively realizing a Two-Higgs-Doublet Model (2HDM) structure with additional heavy states decoupled at the TeV scale. This regime leads to rich phenomenology accessible at current colliders, including modified Higgs couplings, new decay channels, and potential signals in flavor physics. (ii) Conversely, when $f \gtrsim v_{\chi^\prime}$, all new scalars become heavy and degenerate at the 3-3-1 scale. In this case, the low-energy theory reduces to the SM augmented by higher-dimensional operators suppressed by powers of $v_{\chi^\prime}/v_{\text{SM}}$, making direct detection challenging and shifting the focus to precision measurements and indirect searches~\cite{Pinheiro:2022yex}.

This dichotomy in the scalar spectrum illustrates an important feature of the 331RHN model: depending on the symmetry-breaking pattern, it can manifest either as a multi-Higgs model with rich collider phenomenology or as an effective SM with small deviations observable only through precision tests. Determining which regime is realized requires dedicated experimental searches for both direct production of new scalars and precision measurements of SM processes.

We now review the 331RHN gauge sector. It contains the eight gluons, eight SU$(3)_L$ bosons $W^\mu_1,\dots,W^\mu_8$, and a single U$(1)_N$ boson $W^\mu_N$. The SU$(3)_L\times$U$(1)_N$ sector mixes to yield the four SM electroweak bosons and five additional heavy states. To obtain the corresponding spectrum, we expand the scalar kinetic term,
\begin{equation}
\sum_{\Phi=\eta,\rho,\chi}({\cal D}^\mu \Phi)^{\dagger} ({\cal D}_\mu \Phi),
\end{equation}
where the covariant derivative is defined as
\begin{equation}
    {\cal D}_\mu = I\partial_\mu  -igW^a_\mu T^a-ig_N N W^N_\mu I,
\end{equation}
where $T^a= \frac{\lambda^a}{2}$ ($\lambda^a$ being the Gell-mann matrices) with $a=1,2,...,8$, $g$ is the SU$(3)_L$ gauge coupling and $g_N$ is the coupling associated to U$(1)_N$. The matrix $W^a_\mu T^a$ is given by
\begin{equation}
W^a_\mu T^a=\frac{1}{2}
\begin{pmatrix}
W^3_\mu + \frac{1}{\sqrt{3}}W
^8_\mu & W^1_\mu -iW^2_\mu &  W^4_\mu -iW^5_\mu \\
W^1_\mu +iW^2_\mu &- W^3_\mu + \frac{1}{\sqrt{3}}W^8_\mu & W^6_\mu -iW^7_\mu\\
 W^4_\mu +iW^5_\mu & W^6_\mu +iW^7_\mu &  -\frac{2}{\sqrt{3}}W^8_\mu
\end{pmatrix}.
\label{mass_gauge}
\end{equation}
From this kinetic term, after the spontaneous symmetry breaking of $\text{SU}(3)_L \times \text{U}(1)_N$ into the standard $\text{SU}(2)_L \times \text{U}(1)_Y$, we obtain the following set of eigenstates:  
\begin{eqnarray}
    &&W^1_\mu\,\,\,,\,\,\, W^2_\mu\,\,\,,\,\,\, W^{\prime \pm}_\mu=\frac{W^6_\mu \mp iW^7_\mu}{\sqrt{2}}\,\,\,,\,\,\,U^0_\mu=\frac{W^4_\mu-iW^5_\mu}{\sqrt{2}},\nonumber \\
    &&U^{0 \dagger}_\mu=\frac{W^4_\mu+iW^5_\mu}{\sqrt{2}}\,\,\,,\,\,\, W^3_\mu\,\,\,,\,\,\, B_\mu\,\,\,,\,\,\, Z^{\prime}_\mu,
\end{eqnarray}
where 
\begin{equation}
     B_\mu=-\frac{t_W}{\sqrt{3}}W^8_\mu +\sqrt{1-\frac{t^2_W}{3}}W^N_\mu \quad \quad \text{and} \quad \quad Z^{\prime}_\mu=\sqrt{1-\frac{t^2_W}{3}}W^8_\mu + \frac{t_W}{\sqrt{3}}W^N_\mu\,,
\end{equation}
and $t_W = \tan \theta_W$, with $\theta_W$ being the Weinberg angle. The neutral gauge boson $B_\mu$ is associated with the hypercharge generator of the U$(1)_Y$ abelian group.
Then, after the electroweak symmetry breaking of $\text{SU}(2)_L \times U(1)_Y$ into $\text{U}(1)_{\text{em}}$, the $W^1_\mu$ boson mixes with $W^2_\mu$ to form the  standard physical charged gauge bosons, $W^{\pm}_\mu=\frac{W^1_\mu \mp iW^2_\mu}{\sqrt{2}}$, and $W^3_\mu$ combines with $B_\mu$ to compose the photon and the standard massive neutral gauge boson, $A_\mu =s_W W^3_\mu +c_W B_\mu$ and $Z_\mu=c_W W^3_\mu -s_W B_\mu$, where $s_W = \sin \theta_W$ and $c_W = \cos \theta_W$. 

In summary, the 331RHN model contains the following spectrum of gauge bosons: 
\begin{equation}
W^{\pm}_\mu\,\,\,,\,\,\, W^{\prime \pm }_\mu\,\,\,,\,\,\,U^0_\mu \,\,\,,\,\,\, U^{0 \dagger}_\mu \,\,\,,\,\,\, A_\mu \,\,\,,\,\,\,Z_\mu\,\,\,,\,\,\,Z^{\prime}_\mu,
\end{equation}
whose masses correspond to
\begin{eqnarray}
    && m^2_W=\frac{g^2}{4}(v^2_\eta + v^2_\rho)\,\,\,,\,\,\, m^2_{W^{\prime}}=\frac{g^2}{4}(v^2_\eta + v^2_{\chi^{\prime}})\,\,\,,\,\,\,m^2_{U^0}=\frac{g^2}{4}(v^2_\rho + v^2_{\chi^{\prime}}),\nonumber \\
    &&  m^2_Z=\frac{g^2}{4c^2_W}(v^2_\eta + v^2_\rho)\,\,\,,\,\,\,  m^2_{Z^{\prime}}=\frac{g^2}{3-4s^2_W}  v^2_{\chi^{\prime}} \,\,\,,\,\,\, m_A^2 =0\,\,\,.
    \label{GBmass}
\end{eqnarray}
From these expression, we infer that
\begin{equation}
    v^2_\eta + v^2_\rho = v_\text{SM}^2\,,
    \label{recoveringSMvev}
\end{equation}
with $v_\text{SM} = 246$ GeV. Also, to match the 331RHN model with the SM, it can be shown that $g$ and $g_N$ are related by
\begin{equation}
    \frac{g}{g_N}=\frac{3\sqrt{2}\sin \theta_W (m_{Z^\prime}) }{\sqrt{3-4\sin^2 \theta_W  (m_{Z^\prime})}},
\end{equation}
where $\sin \theta_W(m_{Z^\prime})$ denotes the weak mixing angle evaluated at the $Z^\prime$ scale.

A notable feature of the 331RHN model is that $\sin\theta_W(m_{Z^\prime}) < \tfrac{3}{4}$. For a detailed discussion on the value of $\sin\theta_W$ in 3-3-1 models, see Ref.~\cite{Buras:2014yna}. Note also that $Z_\mu$ and $Z^\prime_\mu$ mix into two physical states, $Z_1 = Z \cos \theta_{331} - Z^\prime \sin \theta_{331}$ and $Z_2 = Z \sin \theta_{331} + Z^\prime \cos \theta_{331}$, where $\sin \theta_{331}$ is given by~\cite{Buras:2014yna}
\begin{equation}
   \sin \theta_{331} = -\frac{m_Z^2 \left(s_W^2 \left(v_\eta ^2+v_\rho ^2\right)^2- c_W^2 \left(v_\eta^2-v_\rho^2\right)^2\right)}{m_{Z^\prime}^2 \left(v_\eta^2+v_\rho^2\right)^2 \sqrt{3-4 s_W^2} }\,.
\end{equation}
This mixing angle is strongly suppressed and can be neglected in the gauge boson mass spectrum~\cite{Long:2003pt,Pires:2006bj,Cogollo:2007qx,Buras:2014yna}, although it plays a role in flavor-changing neutral current processes, such as meson transitions, as discussed in Section~\ref{sec:state_of_art}. 

Moreover, for $v_{\chi^\prime} \gg v_\eta, v_\rho$, it follows that
\begin{equation}
    m_{W'^+} \approx m_{U^0} \approx \frac{\sqrt{3-4\sin^2\theta_W}}{2} m_{Z^\prime} \approx 0.72\, m_{Z^\prime},
\end{equation}
implying that $Z^\prime$ is the heaviest gauge boson in the model.
\section{The structure of flavor-changing neutral interactions in the 331RHN}\label{sec:structure}

As mentioned previously, FCNC processes in 3-3-1 models are unavoidable and arise at tree level, a stark contrast to the SM, where such processes are forbidden by the GIM mechanism and appear only at loop level, suppressed by the smallness of CKM entries and loop factors. The tree-level emergence of FCNC in 3-3-1 models has a fundamental origin: anomaly cancellation requires that one quark family transforms as a triplet under SU$(3)_L$ while the other two transform as anti-triplets. This non-universal transformation property breaks the flavor symmetry that protects the SM from tree-level FCNC. Consequently, the neutral gauge boson $Z^\prime$ and the neutral scalars ($h$, $H$, $A$, $H^\prime$) couple non-diagonally to quarks in the mass basis, mediating FCNC processes such as $K$--$\bar{K}$ transitions, $B_s$--$\bar{B}_s$ transitions, and rare decays, e.g., $b \to s \gamma$ and $B_s \to \mu^+ \mu^-$. 
Since the mass of $H'$ is proportional to $v_{\chi'}$, it is one of the heaviest neutral scalars, and its contribution is usually neglected in FCNC analyses. The standard $Z$ boson contributes to these processes only through $Z$-$Z^\prime$ mixing, which is suppressed by $(v_{\text{SM}}/v_{\chi'})^2$ and hence negligible for $v_{\chi^\prime} \gtrsim $ a few TeV~\cite{Long:2003pt,Pires:2006bj,Cogollo:2007qx}. This tree-level structure makes FCNC observables extraordinarily sensitive probes of the 3-3-1 scale, often providing constraints comparable to or stronger than direct searches at colliders.

Before discussing FCNC in detail, we establish the notation for quark mass eigenstates. The quark fields in the interaction (flavor) basis, $u_{L,R}$ and $d_{L,R}$, are related to the physical mass eigenstates, $\hat{u}_{L,R}$ and $\hat{d}_{L,R}$, by unitary transformations:
\begin{equation}
 \hat{u}_{L,R}=V^{\dagger u}_{L,R} u_{L,R}\,\,\,,\,\,\, \hat{d}_{L,R}=V^{\dagger d}_{L,R} d_{L,R}.   
\end{equation}
These mixing matrices generalize the CKM matrix of the SM and encode the non-trivial flavor structure induced by the extended gauge symmetry. The mismatch between left- and right-handed rotations, $V^u_L \neq V^u_R$ and $V^d_L \neq V^d_R$, is precisely what generates tree-level flavor changing couplings to neutral currents, a phenomenon absent in the SM where $V^u_R$ and $V^d_R$ can be absorbed into field redefinitions without physical consequences.

Throughout this work, we use a basis where the right-handed quark fields are diagonal. In this convention, all flavor mixing is encoded in the left-handed rotation matrices $V_{L}^{u}$ and $V_{L}^{d}$, which satisfy the relation $V^{u}_{L} V^{d \dagger}_{L}=V_\text{CKM}$, where $V_\text{CKM}$ is the CKM matrix~\cite{PDG2024}:
\begin{equation}|V_{\text{CKM}}|=
\begin{pmatrix}
0.97435 \pm 0.00016 & 0.22501 \pm 0.00068 & 0.003732^{+0.000090}_{-0.000085} \\
0.22487 \pm 0.00068 &  0.97349 \pm 0.00016 &  0.04183^{+0.00079}_{-0.00069} \\
0.00858^{+0.00019}_{-0.00017} & 0.04111^{+0.00077}_{-0.00068} & 0.999118^{+0.000029}_{-0.000034}
\end{pmatrix}.
\label{eq:CKM_values}
\end{equation}

We now focus on Yukawa interactions. The fact that one fermion family transforms as a triplet and the others as anti-triplets under SU$(3)_L$ requires fermion masses to arise from more than one VEV, leading to FCNC. Depending on which family transforms differently, three distinct sets of Yukawa interactions can be written, each defining a variant of the model. Thus, each 3-3-1 model has three variants. Below, we present these three variants of the 331RHN model in a general form.

Variant I corresponds to the scenario where the first two quark families transform as anti-triplets and the third family as a triplet under $\text{SU}(3)_L$. In this case, the quark content and its transformations under the gauge symmetry are:
\begin{eqnarray}
&&Q_{i_L} = \left (
\begin{array}{c}
d_{i} \\
-u_{i} \\
d^{\prime}_{i}
\end{array}
\right )_L\sim(3\,,\,\bar{3}\,,\,0)\,,u_{iR}\,\sim(3,1,2/3),\,\,\,\nonumber \\
&&\,\,d_{iR}\,\sim(3,1,-1/3)\,,\,\,\,\, d^{\prime}_{iR}\,\sim(3,1,-1/3),\nonumber \\
&&Q_{3L} = \left (
\begin{array}{c}
u_{3} \\
d_{3} \\
u^{\prime}_{3}
\end{array}
\right )_L\sim(3\,,\,3\,,\,1/3),u_{3R}\,\sim(3,1,2/3),\nonumber \\
&&\,\,d_{3R}\,\sim(3,1,-1/3)\,,\,u^{\prime}_{3R}\,\sim(3,1,2/3),
\label{quarks_VariantI} 
\end{eqnarray}
where  the index $i=1,2$ is restricted to only two generations. The negative sign in the anti-triplet $Q_{i_L}$ is used to standardize the signs in charged-current interactions with the gauge bosons.  The primed quarks are new heavy quarks with the usual $(+\frac{2}{3}, -\frac{1}{3})$ electric charges. The simplest Yukawa interactions that generate the correct masses for all standard quarks and are invariant under the $Z_2$ symmetry are given by the terms\footnote{For the most general Yukawa interactions involving lepton-number-violating terms, see:~\cite{Doff:2006rt}.},
\begin{equation}\label{yukawa1}
-{\cal L}_Y \supset g^1_{ia} \bar Q_{i_L} \eta^* d_{a_R} + h^1_{3a} \bar Q_{3_L} \eta u_{a_R} + g^1_{3a} \bar Q_{3_L} \rho d_{a_R} + h^1_{ia} \bar Q_{i_L} \rho^* u_{a_R} + \mbox{H.c.}\,,
\end{equation}
where $a=1,2,3$ and the parameters $g^1_{ab}$ and $h^1_{ab}$ are Yukawa couplings that, for the sake of simplification,  we assume to be real.

In what we call variant II, the second family transforms as a triplet, while the first and third transform as anti-triplets. The quark content transforms as
\begin{eqnarray}
&&Q_{2_L} = \left (
\begin{array}{c}
u_{2} \\
d_{2} \\
u^{\prime}_{2}
\end{array}
\right )_L\sim(3\,,\,3\,,\,1/3)\,,u_{2R}\,\sim(3,1,2/3),\,\,\,\nonumber \\
&&\,\,d_{2R}\,\sim(3,1,-1/3)\,,\,\,\,\, u^{\prime}_{2R}\,\sim(3,1,2/3),\nonumber \\
&&Q_{iL} = \left (
\begin{array}{c}
d_{i} \\
-u_{i} \\
d^{\prime}_{i}
\end{array}
\right )_L\sim(3\,,\,\bar 3\,,\,0),u_{iR}\,\sim(3,1,2/3),\nonumber \\
&&\,\,d_{iR}\,\sim(3,1,-1/3)\,,\,d^{\prime}_{iR}\,\sim(3,1,-1/3),
\label{quarks_VariantII} 
\end{eqnarray}
where $i=1,3$. The Yukawa interactions among the scalars and the standard quarks satisfying the $Z_2$ symmetry are
\begin{eqnarray}
&-&{\cal L}_Y \supset g^2_{2a}\bar Q_{2_L}\rho d_{a_R}+ g^2_{ia}\bar Q_{i_L}\eta^* d_{a_R} 
+h^2_{2a} \bar Q_{2_L}\eta u_{a_R} +h^2_{ia}\bar Q_{i_L}\rho^* u_{a_R} + \mbox{H.c.}\,.
\label{yukawa2}
\end{eqnarray}

Finally, variant III is characterized by the first family transforming as triplet and the second and third as anti-triplets, so the quark content is defined by
\begin{eqnarray}
&&Q_{1_L} = \left (
\begin{array}{c}
u_{1} \\
d_{1} \\
u^{\prime}_{1}
\end{array}
\right )_L\sim(3\,,\,3\,,\,1/3)\,,u_{1R}\,\sim(3,1,2/3),\,\,\,\nonumber \\
&&\,\,d_{1R}\,\sim(3,1,-1/3)\,,\,\,\,\, u^{\prime}_{1R}\,\sim(3,1,2/3),\nonumber \\
&&Q_{iL} = \left (
\begin{array}{c}
d_{i} \\
-u_{i} \\
d^{\prime}_{i}
\end{array}
\right )_L\sim(3\,,\,\bar 3\,,\,0),u_{iR}\,\sim(3,1,2/3),\nonumber \\
&&\,\,d_{iR}\,\sim(3,1,-1/3)\,,\,d^{\prime}_{iR}\,\sim(3,1,-1/3),
\label{quarks_VariantIII} 
\end{eqnarray}
with $i=2,3$, and the respective Yukawa interactions are given by
\begin{eqnarray}
&-&{\cal L}_Y \supset  g^3_{1a}\bar Q_{1_L}\rho d_{a_R}+ g^3_{ia}\bar Q_{i_L}\eta^* d_{a_R} 
+h^3_{1a} \bar Q_{1_L}\eta u_{a_R} +h^3_{ia}\bar Q_{i_L}\rho^* u_{a_R} + \mbox{H.c.}\,.
\label{yukawa3}
\end{eqnarray}

The FCNC mediated by neutral gauge bosons arises from the fermionic kinetic terms:
\begin{eqnarray}
    \bar f_{L,R}\gamma^\mu{\cal D }_{\mu} f_{L,R},  
\end{eqnarray}
where $f$ represents a fermion of the model. The covariant derivative, ${\cal D}$, acts differently in quark triplets or anti-triplets, so the interactions from these terms are variant-dependent. 

For the case of variant I the covariant derivative acts as
\begin{eqnarray}
 && {\cal D }_{\mu} Q_{i_L}=(\partial_\mu -ig \vec{T}\cdot\vec{W_\mu}-ig_NN(Q_{i_L})W^N_\mu)Q_{i_L}, \nonumber \\
 && {\cal D }_{\mu} Q_{3_L}=(\partial_\mu +ig \vec{T^*}\cdot\vec{W_\mu}-ig_NN(Q_{3_L})W^N_\mu)Q_{3_L},
\end{eqnarray}
with $i=1,2$. 

For variant II the covariant derivative is defined as
\begin{eqnarray}
 && {\cal D }_{\mu} Q_{i_L}=(\partial_\mu -ig \vec{T}\cdot\vec{W_\mu}-ig_NN(Q_{i_L})W^N_\mu)Q_{i_L}, \nonumber \\
 && {\cal D }_{\mu} Q_{2_L}=(\partial_\mu +ig \vec{T^*}\cdot\vec{W_\mu}-ig_NN(Q_{2_L})W^N_\mu)Q_{2_L},
\end{eqnarray}
with $i=1,3$.

Finally, the covariant derivative of variant III is
\begin{eqnarray}
 && {\cal D }_{\mu} Q_{i_L}=(\partial_\mu -ig \vec{T}\cdot\vec{W_\mu}-ig_NN(Q_{i_L})W^N_\mu)Q_{i_L}, \nonumber \\
 && {\cal D }_{\mu} Q_{1_L}=(\partial_\mu +ig \vec{T^*}\cdot\vec{W_\mu}-ig_NN(Q_{1_L})W^N_\mu)Q_{1_L},
\end{eqnarray}
with $i=2,3$.

As mentioned, FCNC processes are an unavoidable feature of 3-3-1 models, arising from the requirement of anomaly cancellation. The flavor structure of the model depends critically on which quark family transforms differently under the $\text{SU}(3)_L$ gauge group. Recent studies indicate that it is the third family that plays this distinct role~\cite{Long:1999ij,Oliveira:2022vjo,Escalona:2025rxu}. This choice is further motivated by phenomenological considerations: the top quark, being the heaviest and shortest-lived quark, decays before it can form mesonic bound states, so does not participate in neutral meson transitions. Then, the only relevant neutral bound state formed by up-like quarks is the $D$ meson---unlike down-type quarks which form $K$, $B_d$, and $B_s$ mesons. Consequently, variant I, where the third family transforms as a triplet, naturally suppresses the most stringent FCNC constraints while maintaining the rich phenomenology in all neutral meson systems. For these reasons, throughout the remainder of this work, we focus mainly on variant I.
\section{FCNC and the  variants of the model}\label{sec:variants}

Having established the general structure of flavor-changing interactions in Section ~\ref{sec:structure}, we now present explicit formulae for the mass differences  the three variants of the 331RHN model. This section provides the complete set of FCNC interactions mediated by neutral scalars ($h$, $H$, $A$) and gauge bosons ($Z,Z^\prime$), highlighting how the choice of variant---determined by which quark family transforms as a triplet under $\text{SU}(3)_L$---directly impacts the flavor structure of these couplings.

Using the scalar spectrum from Section ~\ref{sec:essence} and the Yukawa interactions from Section ~\ref{sec:structure}, we derive the FCNC couplings of neutral scalars to quarks. The set of  interactions mediated by the SM-like Higgs $h$ and the heavier CP-even scalar $H$ that lead to FCNC processes is given by
\begin{eqnarray}\label{eq:FCNC_Yukawa_h}
{\cal L}_Y^{h}&\supset&  \frac{2 }{|\sin(2\phi)| }    \left[  \cos(\phi+\varphi) (V^u_L)_{xa}^* (V^u_L)_{bx} \right]\frac{(m_{\text{up}})_a}{v} \bar{\hat{u}}_{b_L} \hat u_{a_R}h  \\ \nonumber
&-&\frac{2 }{|\sin(2\phi)| }   \left[ \cos(\phi+\varphi)   (V^d_L)_{xa}^* (V^d_L)_{bx} \right]\frac{( m_{\text{down}})_a}{v} \bar{\hat{d}}_{b_L} \hat d_{a_R}h + \text{H.c.},    \end{eqnarray}
\begin{eqnarray}
\label{eq:FCNC_Yukawa_H}
 {\cal L}^{H}_Y&&\supset -2  \frac{\cos (\phi +\varphi)}{|\sin 2\phi|}(V^d_L)_{xa}^* (V^d_L)_{bx} \frac{(m_{\text{up}})_a}{v} \bar{\hat{u}}_{b_L} \hat u_{a_R}H  \\ \nonumber
 &&+  2    \frac{\cos (\phi -\varphi)}{|\sin 2\phi|} (V^d_L)_{xa}^* (V^d_L)_{bx} \frac{(m_{\text{down}})_a}{v} \bar{\hat{d}}_{b_L} \hat d_{a_R}H  + \text{H.c.}\,, \end{eqnarray}
while the interactions involving the CP-odd pseudoscalar $A$  are given by
\begin{eqnarray}
 \label{eq:FCNC_Yukawa_A}
  {\cal L}_Y^{A}&\supset&i\left(\tan{\phi} +  \cot{\phi}     \right)(V^u_L)_{xa}^* (V^u_L)_{bx} \frac{(m_{\text{up}})_a}{v}  \bar{\hat{u}}_{b_L}\hat u_{a_R} A \\
&+& i\left(\tan{\phi}  +\cot{\phi} \right)(V^d_L)_{xa}^* (V^d_L)_{bx}\frac{(m_{\text{down}})_a}{v}  \bar{ \hat{d}}_{b_L}\hat d_{a_R} A  + \text{H.c.} \nonumber\,,
\end{eqnarray}
where $a,b=1,2,3$, with $\hat u_{1,2,3}=(u,c,t)$ and $\hat d_{1,2,3}=(d,s,b)$, and $\varphi$ and $\phi$ are the mixing angles of the CP-even and CP-odd low-energy scalars, respectively, as presented in Section ~\ref{sec:essence}. The angle $\phi$ satisfy $\tan \phi = v_\rho/v_\eta$. Crucially, the relevant entries of the quark mixing matrices in these interactions are parametrized by the index $x$, whose value depends on the variant choice as specified in Tab.~\ref{tab:x_parameters}.
\begin{table}[ht!]
\centering
\begin{tabular}{|c| c |}
\hline
Variant& $\quad x \quad $   \\
\hline
I &   $3$ \\
\hline
II &   $2$ \\
\hline
III &  $1$\\
\hline
\end{tabular}
\caption{The relationship between the index $x$ and the variant choice.}
\label{tab:x_parameters}
\end{table}

It can be seen in Eqs.~\eqref{eq:FCNC_Yukawa_h}--\eqref{eq:FCNC_Yukawa_A} that the scalar-mediated  FCNC amplitudes depend on the mixing angles $\phi$ and $\varphi$, as well as on the quark mixing matrices $V^{u,d}_L$. While the masses of $H$ and $A$ remain free parameters of the model, the SM-like Higgs mass is fixed at $m_h = 125.11$ GeV~\cite{ATLAS:2012yve,CMS:2012qbp}. Being the lightest neutral scalar, $h$ generically provides the dominant contribution to FCNC observables, thereby imposing stringent constraints on the parameter space. These constraints can be significantly relaxed by imposing the \textit{alignment condition}, $\phi + \varphi = \pi/2$, which ensures that the SM-like Higgs couples to quarks in a flavor-diagonal manner at tree level. Under alignment, FCNC processes are mediated exclusively by the heavier scalars $H$ and $A$, whose contributions are expected to be mass-suppressed.

In addition to the scalar contributions discussed above, the $Z^\prime$ gauge boson mediates FCNC at tree level through its non-diagonal couplings to quarks. From the fermionic kinetic terms presented in Section ~\ref{sec:structure}, these flavor-violating interactions are given by\footnote{For a detailed discussion of flavor-diagonal interactions of the $Z$ and $Z^\prime$ bosons with fermions, as well as electromagnetic and charged-current interactions, see~\cite{Long:1995ctv}.}:
\begin{eqnarray}
{\cal L}^{Z^{\prime}}_u&\supset&\frac{g}{c_W} \bar{\hat{u}}_{b_L}\left[ \frac{1-s^2_W}{\sqrt{3-4s^2_W}} (V^u_L)^*_{bx}(V^u_L)_{xa}\right]\gamma^\mu \hat u_{a_L}Z^{\prime}, \label{eq:FCNCZprimeu}  \\
{\cal L}^{Z^{\prime}}_d&\supset&\frac{g}{c_W} \bar{\hat{d}}_{b_L}\left[ \frac{1-s^2_W}{\sqrt{3-4s^2_W}} (V^d_L)^*_{bx}(V^d_L)_{xa}\right]\gamma^\mu \hat d_{a_L}Z^{\prime}\,.
\label{eq:FCNCZprimed}
\end{eqnarray}

The flavor-changing interactions in Eqs.~\ref{eq:FCNC_Yukawa_h}--\ref{eq:FCNCZprimed}, imply the following contributions to the mass difference $\Delta M_M$, governing the transition frequency of the neutral meson $M(\bar{q}^b q^a)$:
\begin{eqnarray}
&\Delta M_{M}^{h} = 8\Bigg(\frac{ \cos(\phi+\varphi)}{\sin  2\phi}\Bigg)^2
   \frac{m_{M} B_{M} f_{M}^2}{m_{h}^2}
   \left[\frac{5}{24}\operatorname{Re}\left[\left(C_{M,h}^L\right)^2+\left(C_{M,h}^R\right)^2\right]
   \left(\frac{m_{M}}{m_{q_a}+m_{q_b}}\right)^2 \right. \nonumber \\
&\quad + \left(2\operatorname{Re}\left[C_{M,h}^L C_{M,h}^R\right]
   \left(\frac{1}{24} + \frac{1}{4}\left(\frac{m_{M}}{m_{q_a}+m_{q_b}}\right)^2\right)\right],
\label{contribution_h} \\
&\Delta M_{M}^{A} = -\frac{2 m_{M} B_{M} f_{M}^2}{m_{A}^2}
   \left[\frac{5}{24}\operatorname{Re}\left[\left(C_{M,A}^L\right)^2+\left(C_{M,A}^R\right)^2\right]
   \left(\frac{m_{M}}{m_{q_a}+m_{q_b}}\right)^2 \right. \nonumber \\
&\quad + \left(2\operatorname{Re}\left[C_{M,A}^L C_{M,A}^R\right]
   \left(\frac{1}{24} + \frac{1}{4}\left(\frac{m_{M}}{m_{q_a}+m_{q_b}}\right)^2\right)\right], \label{contribution_A} \\
&\Delta M_{M}^{H} = \frac{8}{\sin^2(2\phi)}\frac{m_{M} B_{M} f_{M}^2}{m_{H}^2}
   \left[\frac{5}{24}\operatorname{Re}\left[\left(C_{M,H}^L\right)^2+\left(C_{M,H}^R\right)^2\right]
   \left(\frac{m_{M}}{m_{q_a}+m_{q_b}}\right)^2 \right. \nonumber \\
&\quad + \left(2\operatorname{Re}\left[C_{M,H}^L C_{M,H}^R\right]
   \left(\frac{1}{24} + \frac{1}{4}\left(\frac{m_{M}}{m_{q_a}+m_{q_b}}\right)^2\right)\right], \label{contribution_H} \\
&\Delta M_{M}^{Z^\prime} = \frac{4\sqrt{2}G_F c_W^4}{3-4s_W^2}
   \left| \left(V_L^{u,d}\right)^*_{bx}\left(V_L^{u,d}\right)_{xa} \right|^2
   \frac{m_Z^2}{m_{Z^\prime}^2} f_{M}^2 B_{M} m_{M} \times r_{\mathrm{RG}}, \label{contribution_Zprime}
\end{eqnarray}
where $m_Z = 91.2$ GeV is the SM-like $Z$ boson mass and $G_F = 1.166 \times 10^{-5}$ GeV$^{-2}$ is the Fermi constant. The Wilson coefficients $C_{M}^{L,R}$ are written in detail in Appendix \ref{app:coefficients}, and their calculation follows the methodology of Ref.~\cite{Gabbiani:1996hi}.
The hadronic parameters entering these expressions are summarized in Tab.~\ref{tab:parameters_table}. The factor $r_\text{RG}$ accounts for renormalization group (RG) evolution from the scale of new physics down to the meson mass scale; at leading order, $r_\text{RG} = 1$. For details on these factor see  the Appendix \ref{app:RGcorr}. The contribution of the SM-like $Z$ boson, in case of non-negligible mixing with the $Z^\prime$, can be computed replacing $\frac{m_Z}{m_{Z^\prime}}\to \sin \theta_{331}$ in Eq.~\ref{contribution_Zprime}.

\begin{table}[ht!]
\centering
\begin{tabular}{|c|c|c|}
\hline
Meson (M)& $m_M$ [MeV] & $\sqrt{B_M}f_M$ [MeV] \\
\hline
$D$ &   $1864.84 \pm 0.05$~\cite{PDG2024} & $212.7 \pm 8.3$~\cite{FLAG} \\
\hline
$K$ &   $497.611 \pm 0.013$~\cite{PDG2024} & $156.3 \pm 0.6$~\cite{FLAG} \\
\hline
$B_d$ &   $5279.72 \pm 0.08$~\cite{PDG2024} & $225 \pm 9$~\cite{FLAG} \\
\hline
$B_s$ &  $5366.93 \pm 0.10$~\cite{PDG2024} & $274 \pm 8$~\cite{FLAG} \\
\hline
\end{tabular}
\caption{Parameters for neutral meson transitions.}
\label{tab:parameters_table}
\end{table}

The equations above highlight an important distinction between scalar- and gauge-mediated contributions: scalar contributions depend on both the quark mixing matrices $V^{u,d}_L$ and the scalar mixing angles $\phi$ and $\varphi$, while $Z^\prime$-mediated processes depend on $V^{u,d}_L$ alone. Additionally, the $Z-Z^\prime$ mixing angle, $\theta_{331}$, introduces further FCNC sources, with the physical SM-like boson $Z_1$ acquiring flavor-violating couplings proportional to the mixing angle and specific entries of $V^{u,d}_L$. The constraints from  the CKM factorization, $V^{u \dagger}_L V^d_L = V_{\text{CKM}}$, and unitarity, $V_L^{u,d\dagger} V^{u,d}_L = I$, are not sufficient to determine this quark mixing pattern. In the absence of experimental input, an infinite number of configurations remain possible.

Different phenomenological approaches have been developed to address this ambiguity. Some studies adopt specific parametrizations or textures for $V^{u,d}_L$ inspired by quark mass matrix ansätze, while others explore the full parameter space through numerical scans. A complementary strategy considers limiting cases such as $V_L^u = I$ (restricting $Z^\prime$ FCNC to down-type quarks) or $V_L^d = I$ (restricting them to up-type quarks), or more general configurations where both matrices exhibit non-trivial flavor structure. The treatment of $Z$-$Z^\prime$ mixing also varies: some analyzes include it explicitly to constrain the mixing angle from FCNC data, whereas others work in the limit where this mixing is negligible. This diversity of approaches underscores both the richness of the 331RHN flavor structures and the complementary roles of different observables in probing the model's parameter space.

Having established the theoretical framework for FCNC in the 331RHN model, we now review how these processes have been phenomenologically studied over the past three decades. In Section~\ref{sec:state_of_art}, we survey the evolution of FCNC analyzes in 3-3-1 models, highlighting the different strategies employed to constrain the parameter space, from early $Z^\prime$-dominated studies using specific mixing matrix parameterizations to recent investigations incorporating family discrimination, mixing effects $Z-Z^\prime$ and scalar alignment. This historical perspective contextualizes the evolution of FCNC phenomenology in 3-3-1 models, illustrating how successive strategies have progressively improved our understanding of the parameter space and clarified the role of the SM-like Higgs.
\section{The state of the art of FCNC processes within the 331RHN}\label{sec:state_of_art}

The investigation of FCNCs in the 331RHN model has evolved considerably over the past three decades, developing from early $Z^\prime$-dominated analyses to a more comprehensive framework that acknowledges the interplay between the gauge and scalar sectors. This section briefly traces the historical development of FCNC phenomenology in 331RHN models, organizing the discussion chronologically to highlight how theoretical understanding has deepened---from the initial studies that entirely neglected scalar contributions, through the subsequent recognition of the heavy neutral scalars $H$ and $A$, to the recent realization that the SM-like Higgs $h$ plays a decisive role in constraining the model's parameter space. A central theme has emerged: any parametrization of the quark mixing matrices $V_L^{u,d}$ must be validated not only against $Z^\prime$-mediated processes but also against scalar-mediated contributions, particularly those involving the lightest scalar state. The alignment limit---a configuration where the SM-like Higgs decouples from FCNC---marks a crucial development that profoundly reshapes our understanding of the viable parameter space in these models.

Early studies of FCNCs in the 331RHN model appeared in the late 1990s. The first, Ref.~\cite{Long1:1999lv}, examined the variant I of the model. In that work, the authors---neglecting scalar  contributions---examined the impact of $Z^{\prime}$ on the mass differences in neutral meson systems such as  $K-\bar{K}$, $D-\bar{D}$ and $B_d-\bar{B}_d$. The analysis was carried out under the assumption of the Fritzsch ansatz for quark mixing, a pioneering framework proposed to explain the structure of the CKM matrix~\cite{Fritzsch:1978hf}
\begin{equation}
    V^{D}_{ij} \sim \left(\frac{m_i}{m_j}\right)^{1/2} \,,
    \label{fritz}
\end{equation}
\noindent where the  index $i < j$, for $i,j=1,2,3$. As a result,  the  $K-\bar{K}$ mixing yielded a lower bound $m_{Z^{\prime}} > 1.02$ TeV.  Subsequently, these results were refined using the then-available experimental data on the rare kaon decay $K^{+} \to \pi^{+}\nu\bar{\nu}$, which raised the lower bound on $m_{Z^{\prime}}$ to $2.3$--$4.3$ TeV~\cite{Long2:2001lv}.

The constraints obtained under the Fritzsch parametrization, while offering valuable early benchmarks, must be interpreted with caution. As discussed below, any parametrization of the mixing matrices $V^{u,d}_L$ is experimentally viable only if the SM-like Higgs contribution to FCNC processes remains within experimental limits---a requirement not accounted for in these early analyses.

Several papers on FCNC in the 331RHN model appeared in the early 2000s. One of them was Ref.~\cite{Sher:2004Sr}, which focused exclusively on FCNC processes mediated by the $Z^{\prime}$. In addition to neutral meson transitions, the authors studied $B \to K l^+ l^-$, $B \to  \mu^+ \mu^-$, and $B_s \to  \mu^+ \mu^-$ decays. In all cases, the authors found that the contributions from rare $B$ decays are much smaller than those from $B - \bar B$ and $B_s - \bar B_s$ mass differences, and thus the model  predicts no substantial contribution to these rare $B$-decays. Moreover, lower bounds on $m_{Z^{\prime}}$ of order tens of TeV  are obtained if a Fritzsch-like structure for the mixing angles is assumed. 

A similar approach was adopted in Ref.~\cite{Ochoa:2008Mo}; this time employing textures for the quark mass matrices. Specifically, four different ansätze for these matrices were analyzed. By comparing the theoretical predictions with the measured mass difference of the $B_s$ system, the authors derived lower bounds on $m_{Z^{\prime}}$ ranging from $1$ to $30$ TeV.

Towards the end of the 2000s, two notable works broadened the study of FCNC processes in the 331RHN model by introducing new tree-level contributions arising from mixing between exotic and standard quarks, as well as mixing between CP-even and CP-odd neutral gauge bosons~\cite{Cabarcas:2010Jm}. The authors analyzed additional effects on meson--antimeson transitions due to the presence of exotic quarks and found consistency with $m_{Z^{\prime}}$ values of order tens of TeV.  Following this direction, the authors of Ref.~\cite{Ponce:2009Wb} derived bounds on $m_{Z^{\prime}}$ from neutral meson mixing and rare quark decay data, examining scenarios where standard quarks mix with exotic ones. A key result was that if the CKM matrix element $V_{tb}$ is significantly smaller than $1$, the model predicts enhanced rates for rare top decays such as $t \to c \gamma$, $t \to c Z$, and $t \to c g$. However, to remain consistent with current experimental constraints, $m_{Z^{\prime}}$ must exceed $10$ TeV provided that $V_{tb} \approx 1$.

Throughout the 1990s and the first decade of the 2000s, the theoretical landscape of FCNCs in 331RHN models shared a common feature: constraints were derived almost exclusively from $Z^\prime$-mediated processes, with each study relying on specific parametrizations of the quark mixing matrices. While this approach yielded valuable early insights, it overlooked a crucial component of the model's phenomenology---the scalar sector, and in particular, the role of the SM-like Higgs boson. Some works in the 2010s continued to examine only $Z^{\prime}$-mediated FCNC processes. For instance, in Ref.~\cite{Queiroz:2016gif}, the authors examined the complementarity between FCNC constraints and dilepton resonance searches at the LHC Run 2, conducted at $\sqrt{s} = 13$ TeV with an integrated luminosity of $3.2$ fb$^{-1}$.  The analysis concluded that a  $Z^{\prime}$ with mass in the few-TeV range remains consistent with FCNC bounds, depending on the chosen parametrization of the quark mixing matrices.

The first study investigating FCNC processes in the 331RHN model mediated by neutral scalars was presented  in the early 2010s in Ref.~\cite{Farinaldo1:2012Df}. In this work, the authors concluded that the most stringent constraint on the model, based on a specific parametrization of the quark mixing matrices $V^{u,d}_L$, arises from the precise measurement of the  $B_d -\bar B_d$ system. This analysis yielded the bounds $m_{Z^{\prime}}>4 $ TeV and $m_H\,,\,m_A>7$ TeV. The inclusion of heavy neutral scalars $H$ and $A$ represented a valuable step forward; however, neglecting the contribution from SM-like Higgs remains as an oversimplification. 

A subsequent work took an opposite approach, neglecting $Z^{\prime}$ contributions to $K -\bar K$ mixing and focusing exclusively on neutral scalar mediators, including the SM-like Higgs~\cite{Okada:2016whh}. The authors showed that, for a specific choice of the $V^{u,d}_L$ parametrization, the heavy scalars $H$ and $A$ can have masses at the electroweak scale while remaining consistent with meson-mixing constraints. 

Still in the 2010s, Buras and his collaborators  initiated a series of investigations on   flavor physics  within the framework of 3-3-1 models. The first one  we highlight here is Ref.~\cite{Buras:2012Rw}\footnote{See also~\cite{Buras:2012jb,Buras:2013rqa}.}. This paper  explored a broad range of observables, including meson--antimeson transitions, rare semileptonic decays, and the top quark decay channel $t \to c \gamma$. All  processes were mediated exclusively by  $Z^{\prime}$ and assumed a specific parametrization of $V^{u,d}_L$. Through a rigorous analysis of $\Delta F=2$ and $\Delta F=1$ processes, they demonstrated that a $Z^{\prime}$ boson with a mass at the TeV scale can still be consistent with existing flavor physics constraints.

The second paper we highlight  is~\cite{Buras:2013dea}. In this paper the authors  investigated how 3-3-1 models confronted new data on $B_{s,d} \to  \mu^+ \mu^-$and $ B_d \to  K^* (K) \mu^+ \mu ^-$ taking into account constraints from $ \Delta F = 2$ observables, low energy precision measurements, LEP-II and LHC data. For the particular case of the 331RHN model, and considering a specific parametrization for the $V^{u,d}_L$, the authors concluded that a $Z^{\prime}$ with mass of a few TeV could accommodate the flavor puzzle at that time.

Another important contribution came later with Ref.~\cite{Buras:2014yna}, where the authors investigated the impact of the $Z-Z^{\prime}$ mixing. They  calculated the impact of this mixing on rare $K$, $B_s$ and $B_d$ decays. For the particular case of the 331RHN, they found that the mixing can indeed be neglected in $\Delta F = 2$ transitions and in decays such as $B_d\to K^* \mu^+ \mu^-$~\cite{LHCb:2013zuf,LHCb:2013ghj}. However, tree-level $Z_1$ exchanges can produce sizable corrections in decays sensitive to axial-vector couplings—such as $B_{s,d} \to \mu^+\mu^-$ and $B_d \to K\mu^+\mu^-$—and in channels involving neutrinos in the final state, like $b \to s\nu\bar\nu$, $K^+ \to \pi^+\nu\bar\nu$, and $K_L \to \pi^0\nu\bar\nu$, which generally cannot be neglected. They also analyzed for the first time the ratio $\epsilon^{\prime}/\epsilon$ in 3-3-1 models including both $Z$ and $Z^\prime$ contributions.  Importantly, they also demonstrated that the choice of fermion representations under $\mathrm{SU}(3)_L$ has direct phenomenological consequences. As the main result, they concluded that  the interplay of new physics effects in electroweak precision observables and flavor observables could help identify the favored 3-3-1 model.

Motivated by the findings that the ratio  $\epsilon^{\prime}/\epsilon$ in the SM appeared significantly below the experimental data,  the paper~\cite{Buras:2015kwd}  investigated whether the necessary enhancement of $\epsilon^{\prime}/\epsilon$ could be obtained in 3-3-1 models. The work took into account contributions from $Z_1$ and $Z_2$. In a subsequent paper~\cite{Buras:2016dxz}, the authors addressed the tension in $\Delta F=2$ processes in relation to $\epsilon^{\prime}/\epsilon$, $B_s \to \mu^+ \mu^-$ and   $B \to  K^* \mu^+ \mu^-$.

Similar methodologies were later adopted in the early 2020s. Notably, in Ref.~\cite{Buras:2021rdg} the authors provided a detailed analysis of the charm sector within 3-3-1 models, while in Ref.~\cite{Colangelo:2021myn}  the decay $B_c \to  B^{* +}\nu \bar \nu$ was analyzed and correlated to the processes $K \to  \pi \nu \bar \nu $ and  $B \to  (X_s, K, K^* )\nu \bar \nu$. Collectively, all these studies reinforced that a $Z^{\prime}$ boson with mass in the TeV range remains compatible with the experimental bounds available at that time. More recently, in the paper~\cite{Buras:2023ldz}, the authors updated the predictions of 3-3-1  models for rare $B$ and $K$ decays, and $\Delta F = 2$ processes.

We highlight here that in all these  works, the authors deliberately omitted the contributions from neutral scalar particles in $\Delta F = 2$ processes. Moreover, it is important to emphasize the strong dependence of $Z^{\prime}$ mass constraints from FCNC processes on the specific parameterizations of the quark mass matrices. This line of investigation persisted as the dominant approach until the early 2020s~\cite{Farinaldo2:2016Fv,Farinaldo3:2023Fc}. In these works, the authors employed ansätze for quark mixing matrix parametrizations inspired by texture-zero structures to examine FCNC effects  mediated by the  $Z^{\prime}$ boson.

To illustrate how different parametrizations of quark mixing, determined from specific configurations of $V^{u,d}_L$, can substantially affect predictions for meson transitions, consider a comparison between Ref.~\cite{Farinaldo1:2012Df} and Ref.~\cite{Farinaldo3:2023Fc}. In the former, a particular parametrization of the $V^{u,d}_L$ matrices led to the bound $m_{Z^{\prime}} > 4$ TeV from the $B_d-\bar B_d$ transition. In contrast, the latter revisited the same process using updated experimental data and an alternative parametrization, obtaining instead $m_{Z^{\prime}} > 400$ TeV. This two-order-of-magnitude discrepancy underscores the sensitivity of FCNC constraints to the assumed quark mixing structure. 

This large variation in constraints---spanning two orders of magnitude---marks a fundamental challenge in FCNC phenomenology within 3-3-1 models: \textit{Which parametrizations, if any, are physically consistent once all mediators are included?} Addressing this problem requires a systematic framework that incorporates all neutral mediators, both scalars and gauge bosons, while simultaneously considering all four neutral meson systems.

As mentioned in previous sections, a key feature of 3-3-1 models is the need to assign one quark family to transform as a triplet under SU$(3)_L$, while the other two transform as anti-triplets, a choice that directly impacts FCNC processes. In particular, Ref.~\cite{Oliveira:2022vjo} analyzed the contribution of the pseudoscalar $A$, which in principle can be even lighter than the SM-like Higgs, to the $K - \bar{K}$ transition. Their results demonstrated that different family assignments lead to distinct bounds on the pseudoscalar mass, with the case where the third family transforms as a triplet (variant I) being the most favored, highlighting the impact of the family assignment on FCNC processes\footnote{This result was further confirmed in a scenario where standard particles mix with exotic  3-3-1 particles~\cite{Huitu:2024nap}. This mixing occurs when the fields $ \eta^{\prime 0}$ and $\chi^0$ acquire non-zero VEVs.}. Results on the issue of family discrimination are presented in detail in Section~\ref{subsec:family}.

The role of the SM-like Higgs boson in FCNC processes was examined in detail in  Ref.~\cite{Oliveira:2022dav}. The terms in Eq.~\ref{eq:FCNC_Yukawa_h} leads to top quark decays such as $t\to h \, c$ and $t \to h \, u$,  through Yukawa interactions,
\begin{equation}
    \mathcal{L}_h^Y \supset \lambda_{tch} \bar t h c + \lambda_{tuh} \bar h_1 u + \text{h.c.}\,.
\end{equation}
The ATLAS bounds on these top decays are found in Ref.~\cite{ATLAS:2017tas}  and impose $\lambda_{tch} < 0.13$
and $\lambda_{tuh} < 0.13 $. In this case, it is necessary to verify if the Yukawa interactions in 
Eq.~\ref{eq:FCNC_Yukawa_h} satisfy such bounds. The authors found both $\lambda_{tch} $ and $\lambda_{tuh}$ to be of order $\sim \mathcal{O}(10^{-4})$.

In addition to the contributions from $h$, $H$, $A$, and $Z^\prime$, the 3-3-1 framework also features $Z$-$Z^\prime$ mixing, which introduces additional sources of FCNC. This mixing, governed by the angle $\theta_{331}$, implies that the SM-like boson $Z$ can also mediate meson transitions. Ref.~\cite{Oliveira:2022dav} analyzed this effect by studying the combined contributions of $h$, $Z^\prime$, and $Z$ to $\Delta M_K$, $\Delta M_{B_d}$, and $\Delta M_D$. By exploring different scenarios for the relative weight of each mediator, they showed that meson transitions serve as a powerful indirect probe of the $Z$–$Z^\prime$ mixing angle, while remaining consistent with collider bounds on flavor-changing top decays. Bounds on this mixing angle are further discussed in Section~\ref{subsec:mixing}.

These studies emphasized that any proposed parametrization of the quark mixing matrices $V^{u,d}_L$ must be validated against the SM-like Higgs contribution to meson-antimeson transition observables, ensuring that its effect remains within the experimental uncertainty of $\Delta M$ measurements.  As an illustrative case, the previous paper proposed a parametrization for  $V^{u,d}_L$ that yields a contribution from  the SM-like Higgs to  $\Delta M_K$ well within the experimental error margins. Once this parametrization was shown to be consistent with the SM-like Higgs contribution,, the analysis proceeded to assess the impact of the $Z^{\prime}$  on  $K -\bar K$ transition. The results revealed that, depending on how significantly the  $Z^{\prime}$-mediated process contributes to the experimental uncertainty in $\Delta M_K$ relative to the scalar sector, the allowed $Z^{\prime}$ mas range extends from a few hundred GeV up to several hundred TeV.

An important step toward a more realistic treatment of flavor physics in the 3-3-1 framework was achieved in Ref.~\cite{Oliveira:2022dav}, where the authors carried out a detailed and analysis of FCNC mediated by both scalar and gauge sectors. In that work, the scalar mass matrix was explicitly constructed and diagonalized, enabling the identification of the SM-like Higgs boson within the 3-3-1 scalar spectrum. To handle the complexity of the scalar potential, the authors adopted the so-called \textit{decoupling limit}, a specific region of parameter space in which the SM-like Higgs reproduces the SM behavior, while the heavy neutral scalars naturally decouple from low-energy flavor observables. In that article, the quark mixing matrices were determined by assigning specific weights to the Higgs contributions, and the resulting impact on $Z'$-mediated transitions was subsequently evaluated.

Another important investigation into the role of the SM-like Higgs in flavor physics within the 331RHN model was recently presented in Ref.~\cite{Escalona:2025rxu}. Unlike $Z^{\prime}$-mediated FCNCs, scalar-induced meson transitions cannot be suppressed by simply increasing the mediator mass~\cite{Cogollo:2012ek,Okada:2016whh,Oliveira:2022vjo}, since one particle must reproduce the 125 GeV boson of the SM. To quantitatively assess the impact of the SM-like Higgs boson on flavor observables, this work formally introduced the concept of the \textit{alignment limit}, which specifies the conditions on scalar mixing that suppress the participation of this particle in FCNC interactions.

It was observed that three crucial features emerge from the Lagrangian in Eq.~\ref{eq:FCNC_Yukawa_h}. First, for variant I, the FCNC interactions arise exclusively through combinations of the elements of the mixing matrix $(V_L^{u,d})_{3a}^*$ and $(V_L^{u,d})_{b3}$. Index 3 reflects this family-specific transformation property. For alternative scenarios (variants II or III), the relevant indices shift accordingly while maintaining analogous matrix element dependencies. Any phenomenologically consistent FCNC study in this context must respect $V_\text{CKM}$ factorization in $V_L^u$ and $V_L^d$, which imposes non-trivial correlations between the four relevant mesons. On the other hand, note that neither this requirement nor unitarity of individual matrices is sufficient to unambiguously define $V_L^u$ and $V_L^d$. However, among the infinite possible quark mixing scenarios, two trivial cases stand out: either $V_L^u$ or $V_L^d$ can be set equal to the identity matrix, leaving the other as the CKM matrix. These cases represent two extreme, yet illustrative scenarios of quark mixing patterns.

Second, the FCNC interaction mediated by $h$ depends on the amplification factor $f(\phi) \equiv 1/|\sin(2\phi)|$. This function reaches its minimum value $f(\pi/4) = 1$ when $\tan\phi = 1$, which corresponds to the maximum mixing ($\phi = \pi/4$). Any deviation from $\tan\phi = 1$ (either $\tan\phi > 1$ or $\tan\phi < 1$) enhances $f(\phi)$, thus amplifying the contributions of FCNC. This establishes $\tan\phi = \frac{v_\eta}{v_\rho} = 1$ as the optimal parameter choice for the suppression of FCNC.

Third, the critical dependence on $\cos(\phi + \varphi)$ reveals a novel alignment mechanism. The condition $\cos(\phi + \varphi) = 0$ completely decouples the SM-like Higgs $h$ process from the FCNC, defining the \textit{alignment limit} of the 3-3-1 model. This work presents the first systematic derivation of this alignment condition within the 3-3-1 framework, establishing
\begin{equation}
    \phi + \varphi = \frac{\pi}{2}
\end{equation}
as the crucial relationship for Higgs-mediated FCNC suppression.

Note that SM predictions for $\Delta M_{M}$ are consistent with the experimental measurements within the $2\sigma$ range. Therefore, any new physics contribution must remain below the corresponding experimental uncertainty, namely $\Delta M_{M}^{331} \leq \delta(\Delta M_{M}^{\text{exp}})$. In the following, $\Delta M_{M}^{331}$ will denote the total contribution to $\Delta M_{M}$ that arises from the 3-3-1 model.

Having established the central role of the alignment limit and the importance of quark mixing parametrization, we now turn to specific phenomenological applications that have increased our understanding of the 331RHN parameter space. We organize these studies thematically, beginning with the question of family discrimination---\textit{which quark family transforms as a triplet under $\text{SU}(3)_L$ ?}---followed by analyses of $Z$-$Z^\prime$ mixing effects and detailed investigations of the scalar sector's impact on meson transitions and its influence on lower bounds on $m_{Z^\prime}$.

\subsection{Family discrimination}\label{subsec:family}

A distinctive feature of 3-3-1 models is that anomaly cancellation requires one quark family to transform as a triplet under SU$(3)_L$, while the other two must transform as anti-triplets. Since theory does not predict which family should take this role, it becomes essential to explore physical processes that can discriminate among these possibilities. Such an arrangement necessarily leads to FCNC interactions at tree level, mediated by neutral scalars and gauge bosons. While early studies focused on the role of the $Z^\prime$ boson as the dominant mediator~\cite{Ng:1992st,Long:1999ij}, Ref.~\cite{Oliveira:2022vjo} revisited the problem by considering  the pseudoscalar $A$, with was argued to possibly be the lightest particle on the 3-3-1 spectrum~\cite{Pinheiro:2022bcs,Cherchiglia:2022zfy}. This motivated an analysis of whether meson transitions, particularly the $K - \bar{K}$ system, are sensitive to the choice of family assignment, which suggested masses on the TeV range.

The authors derived the effective pseudoscalar interactions with quarks for the three possible variants of family assignment and studied their contributions to $\Delta M_K$. Since the theoretical prediction within the SM suffers from QCD uncertainties but remains consistent with experimental data~\cite{Buchalla:1995vs,DiLuzio:2019jyq,DeBruyn:2022zhw}, they adopted a conservative approach and required that the new contribution from the 3-3-1 model should not exceed the experimental uncertainty, $\Delta M_K^{331} < 0.006 \times 10^{-12}$ MeV~\cite{HFLAV:2019otj,PDBook,Chen:2021ftn,LHCb:2013zpr}. To implement this condition, the quark mixing matrices $V^u_L$ and $V^d_L$ were parameterized consistently with the CKM relation $V^u_L V^{d\dagger}_L = V_{\text{CKM}}$ and explored numerically through random sampling of their entries. This allowed for an exhaustive test of possible textures that respect both unitarity and CKM constraints.  As illustrative example, they present the following texture:
\begin{equation} V^d_L = \begin{pmatrix} 0.849036 & 0.17803 & 0.497437 \\ 0.175894 & -0.983055 & 0.0516103\\ 0.498197 & 0.0436771 & -0.865963 \end{pmatrix}\,, V^u_L = \begin{pmatrix} 0.868672 & -0.0475631 & 0.493095 \\ 0.38298 & -0.915136 & 0.125881\\ 0.512263 & 0.00476072 & -0.858815 \end{pmatrix}\,.
\label{EId} 
\end{equation}
Note that in Ref.~\cite{Oliveira:2022vjo}, the authors neglected the CP phase: $\delta =0$.

\begin{figure}[ht]
    \centering
    \includegraphics[width=0.7\linewidth]{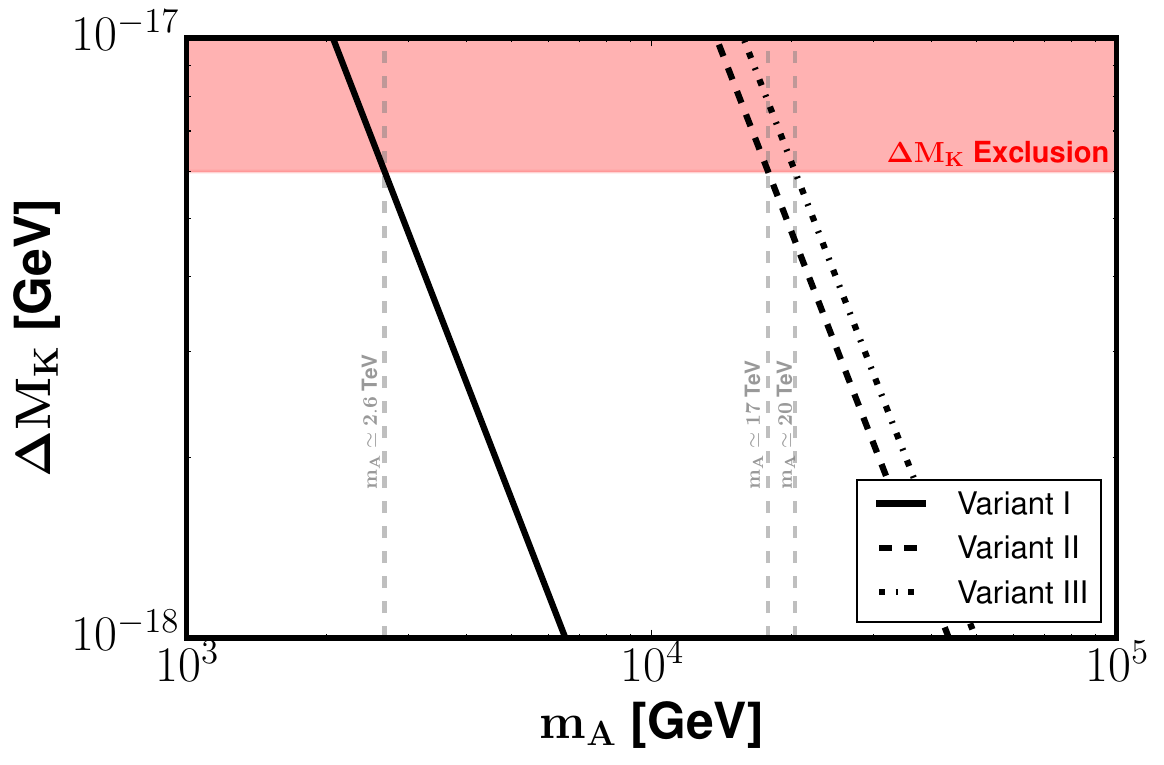}
    \caption{Numerical results for variant I (continuous curve), variant II (dashed curve), and variant III (dashed dotted curve). Figure taken from Ref.~\cite{Oliveira:2022vjo}.}
    \label{fig:family_Disc}
\end{figure}

The numerical analysis revealed that FCNC processes are highly sensitive to family discrimination, as shown in Fig.~\ref{fig:family_Disc}, since the pseudoscalar contribution to $\Delta M_K$ scales with the fourth power of the mixing elements of the matrix, as shown in Eq.~\ref{contribution_A}. Each family assignment resulted in different lower bounds on the pseudoscalar mass. Variants II and III, where the first or second family transforms as a triplet, respectively, produced very stringent bounds, pushing $m_A$ into the multi-TeV regime ($m_A \gtrsim 17$--$20$ TeV). In contrast, variant I, where the third family transforms as a triplet, led to a significantly weaker constraint ($m_A \gtrsim 2.6$ TeV), making it the most energetically favorable configuration. These findings reinforce earlier conclusions that the third family plays a special role in 3-3-1 models~\cite{Ng:1992st,Long:1999ij}, while also establishing the pseudoscalar as a powerful probe of family discrimination. The study thus provided both the most stringent bounds on the pseudoscalar mass in this framework and strong evidence favoring the assignment of the third family to the SU$(3)_L$ triplet representation.

The sensitivity of FCNC processes to family assignment reveals a distinctive feature of 3-3-1 models: flavor observables measured at low energies can directly probe the structure of the gauge symmetry.  The gauge sector, however, offers another way for testing the model. Mixing between the SM $Z$ boson and the exotic $Z^\prime$ introduces additional flavor-changing effects that complement the family discrimination studies.

Other works have also investigated how family discrimination affects FCNC in 3-3-1 models. In particular, Cárcamo, Martínez, and Ochoa~\cite{CarcamoHernandez:2005ka} performed one of the first systematic studies of this connection by analyzing the implications of different fermion embeddings for the neutral current structure. They showed that the choice of which quark family transforms as a triplet or anti-triplet under SU$(3)_L$ leads to distinct patterns of nonuniversal $Z^\prime$ couplings, producing tree-level FCNCs whose size and sign depend sensitively on the family representation. In their analysis, the authors also investigated how family discrimination influences the FCNC effects in the decays of both $Z$ and $Z^\prime$. By confronting the model predictions with precision electroweak data at the $Z$ pole, they demonstrated that $m_{Z^\prime} \lesssim 1.5~\text{TeV}$ is strongly disfavored in the 331RHN, while configurations with heavier $Z^\prime$ remain viable. Among the three possible family assignments, they found that variant III,  leads to the strongest suppression of FCNC effects, making it the most phenomenologically favored configuration. These results, together with later studies such as Ref.~\cite{Oliveira:2022vjo}, established that FCNC observables can serve as a robust probe of family discrimination, tightly linking the gauge representation of quarks to measurable flavor phenomena.

\subsection{\texorpdfstring{$Z$-$Z^\prime$ mixing effects}{Z-Z' mixing effects}}\label{subsec:mixing}

The mixing between $Z$ and $Z^\prime$ is a generic prediction of 3-3-1 models and gives rise to two physical neutral states, $Z_1 = Z \cos \theta_{331} - Z^\prime \sin \theta_{331}$ and $Z_2 = Z \sin \theta_{331} + Z^\prime \cos \theta_{331}$. As a consequence, the SM-like boson $Z_1$ acquires flavor-changing couplings to quarks and contributes to neutral meson transitions. Following Ref.~\cite{Oliveira:2022dav}, the effective Lagrangian was derived for $K - \bar K$, $B_d - \bar B_d$, and $D - \bar D$ transitions, considering the variant I of the model. The $Z$ contribution is proportional to $\sin^2\theta_{331}$ and depend on specific entries of the quark mixing matrices, as described in Section~\ref{sec:variants}. This makes the mixing angle $\theta_{331}$ a directly testable parameter through FCNC observables.  

\begin{figure}
    \centering
    \includegraphics[width=\columnwidth]{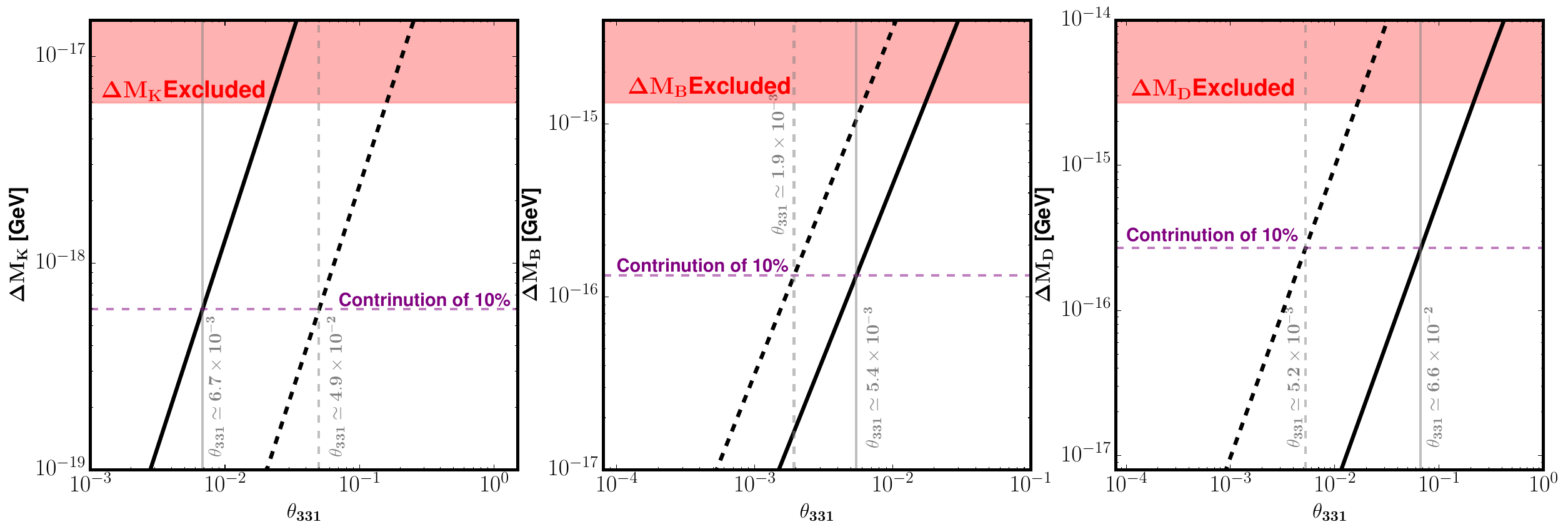}     
    \caption{Evolution of $(\Delta M_K)_Z$, $(\Delta M_B)_Z$, and $(\Delta M_D)_Z$ in function of $\theta_{331}$. The excluded red region represents the error of $\Delta M_{K,B,D}$.  The continuous (dashed) black lines represents the contribution of $Z^\prime$ for $80\%$ ($10\%$) of the error of difference of meson masses. The continuous (dashed) horizontal purple lines represents the contribution of $10\%$ ($80\%$) of $\Delta M_{K,B,D}$.  Figure taken from Ref.~\cite{Oliveira:2022dav}.}
  \label{fig:Z_phi}
\end{figure}

In order to obtain quantitative bounds, the analysis required that the sum of contributions from $h$, $Z^\prime$, and $Z$ respect the experimental uncertainty in the measured mass differences of neutral mesons. Since the relative size of each mediator’s contribution is not fixed by the model, two benchmark scenarios were considered. In the first case, $h$ was assumed to account for $80\%$ of the allowed window, while in the second case only $10\%$ was attributed to $h$. For both scenarios, explicit solutions for the mixing matrices $V^u_L$ and $V^d_L$ were obtained, consistent with CKM unitarity. These textures were then used to evaluate the size of the effective couplings.  

The Fig.~\ref{fig:Z_phi} shows the result obtained in Ref.~\cite{Oliveira:2022dav}. The results show that the $K$, and $B_d$ meson transitions impose the most stringent bounds on $\theta_{331}$, requiring $\theta_{331} \approx 6 \times 10^{-3}$. Such bounds are close to the previous ones found in the literature, see Refs.~\cite{Long:1996rfd,Cogollo:2007qx}. In both benchmark cases, the predicted contributions of $Z_1$ remain within experimental uncertainties, while they agree with the top quark decays constraint ~\cite{ATLAS:2017tas}. 

A complementary  study of $Z$–$Z^\prime$ mixing effects in 3-3-1 models was carried out in Ref.~\cite{Buras:2014yna}. In that work, the authors revisited the general structure of the neutral gauge boson sector for arbitrary values of the $\beta$ parameter and for different fermion representations. They demonstrated that, although $\sin \theta_{331}$ is typically of order $10^{-3}$, the induced flavor-changing couplings of the SM-like boson $Z_1$ can produce non-negligible effects in flavor observables sensitive to axial-vector currents, such as $B_{s,d}\to\mu^+\mu^-$ and $K\to\pi\nu\bar\nu$, while remaining negligible in $\Delta F=2$ transitions. The analysis also clarified that the sign of $\sin \theta_{331}$ and the pattern of interference between $Z$- and $Z^\prime$-mediated amplitudes depend crucially on the choice of fermion representations and on the parameter $\beta$, highlighting the phenomenological differences of 3-3-1 variants. Furthermore, by correlating flavor-changing and electroweak precision observables, they showed that precision measurements of $\sin^2\theta^\ell_{\rm eff}$ and rare meson decays could together discriminate among the viable 3-3-1 constructions. This study established a solid theoretical framework to interpret $Z$–$Z^\prime$ mixing as a key probe of the gauge and scalar sectors of 3-3-1 models.

The analyses discussed until now, family discrimination and $Z$-$Z^\prime$ mixing, represent complementary probes of the gauge structure of 3-3-1 models. We now return to the scalar sector, examining in detail how the SM-like Higgs $h$, together with the heavier states $H$ and $A$, contributes to neutral meson transitions. This discussion condenses the recent theoretical advances that culminated in the alignment limit framework.

\subsection{FCNC and SM-like Higgs}\label{subsec:smlike}

The Yukawa interaction presented in Eq.~\ref{yukawa1} generate FCNC couplings involving neutral scalars once the scalar fields are rotated to their mass eigenstates. As discussed in Section~\ref{sec:essence}, the physical scalar spectrum consists of one massive pseudoscalar $A$ and three CP-even states: $H'$, $H$, and $h$. The heaviest state, $H'$, originates predominantly from the $R_{\chi'}$ component and effectively decouples from the low-energy theory when $v_{\chi'} \gg v_\eta, v_\rho$. The remaining scalars---$A$, $H$, and $h$---form a spectrum characteristic of the 2HDM, where their mixing is governed by the angles $\phi$ and $\varphi$ introduced in Section~\ref{sec:essence}.

The contribution of these scalars to FCNC processes depends on both their masses and the scalar mixing configuration. To isolate the role of the SM-like Higgs $h$ and assess the validity of the alignment limit discussed in Ref.~\cite{Escalona:2025rxu}, we adopt a benchmark scenario where the heavier scalars $A$ and $H$ are sufficiently massive to decouple from low-energy observables. Specifically, we set
\begin{equation}
m_H = m_A = m = 10\ \text{TeV},
\end{equation}
ensuring that their contributions to meson transitions are negligible\footnote{Collider bounds and theoretical consistency require $m_H, m_A \gtrsim 350$ GeV~\cite{Cherchiglia:2022zfy}. Our choice of $10$ TeV represents a conservative scenario where only the SM-like Higgs plays a significant role in FCNC processes.}. The SM-like Higgs mass is fixed at its measured value,
\begin{equation}
m_h = 125.11\ \text{GeV}.
\end{equation}

Under these assumptions, the FCNC phenomenology is controlled by three ingredients: the Higgs mass $m_h$, the scalar mixing angles $\phi$ and $\varphi$, and the quark mixing matrices $V_L^{u,d}$. To systematically explore the relationship between these parameters, we examine three representative scenarios that span the range of possible quark mixing configurations consistent with the CKM factorization $V_L^u V_L^{d\dagger} = V_{\text{CKM}}$.

Each scenario represents a distinct limit of the quark mixing structure and probes different combinations of neutral meson systems:

\begin{itemize}
\item \textbf{Scenario (i)} ($V_L^u = V_{\text{CKM}}^\dagger$, $V_L^d = \mathbf{1}$): FCNC processes are confined to the up-quark sector, with only $D$-$\bar{D}$ mixing providing constraints. This represents the least restrictive case.

\item \textbf{Scenario (ii)} ($V_L^d = V_{\text{CKM}}$, $V_L^u = \mathbf{1}$): FCNC processes occur exclusively in the down-quark sector, being constrained by $K$, $B_d$, and $B_s$ transitions. The precision measurement of $B_s$ mixing makes this the most restrictive scenario.

\item \textbf{Scenario (iii)} (generic $V_L^{u,d}$): A mixed configuration where both up- and down-quark sectors contribute to FCNC, representing an arbitrary intermediate regime.
\end{itemize}

For each scenario, we analyze how the scalar mixing parameters $\phi$ and $\varphi$ are constrained by requiring that the Higgs-mediated contribution $\Delta M_M^h$, Eq.~\ref{contribution_h}, remains within experimental bounds. The alignment condition $\phi + \varphi = \pi/2$ emerges naturally as the parameter configuration that minimizes these contributions.

\begin{figure}[!ht]
    \centering    
    \includegraphics[width=0.49\linewidth]{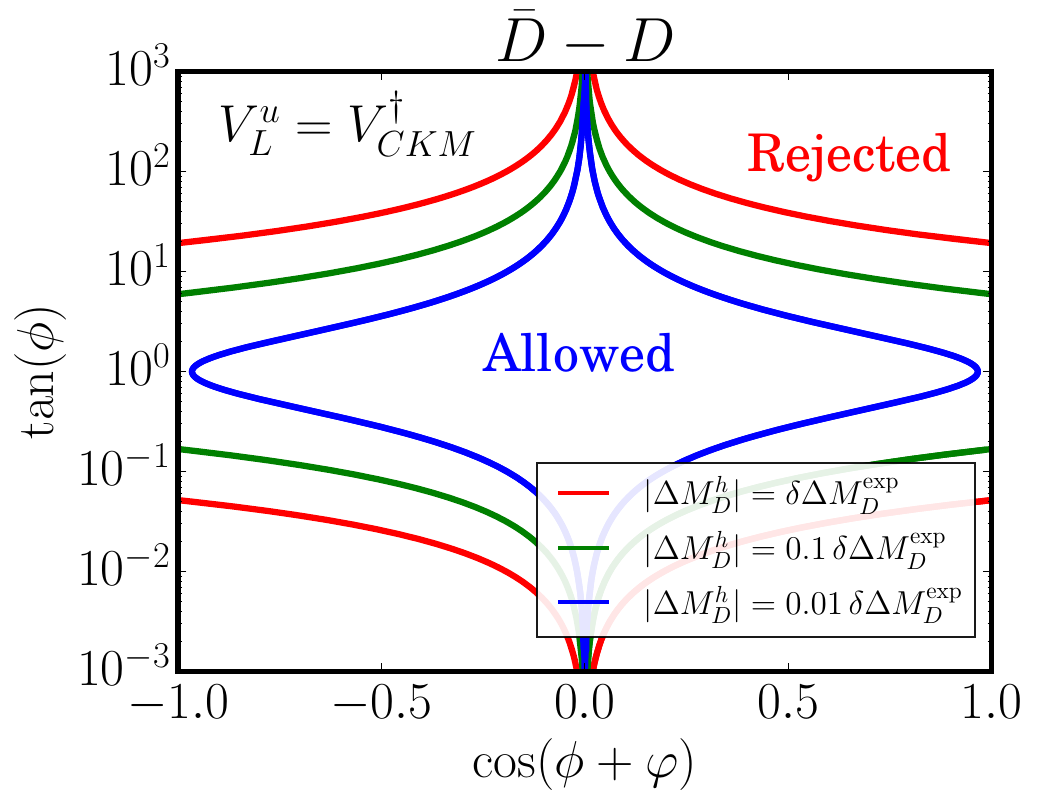}
    \includegraphics[width=0.49\linewidth]{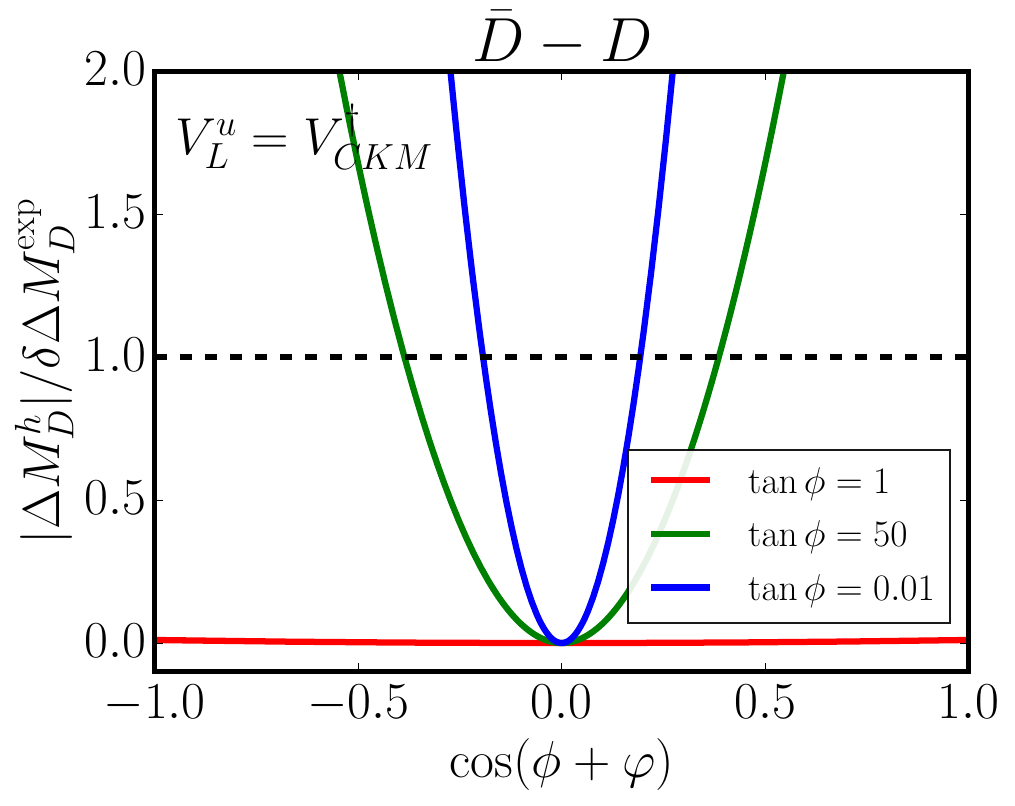}
    \caption{(a) Contours on the $\cos(\phi+\varphi)$--$\tan\phi$ plane fulfilling $100\%$ (red), $10\%$ (green) and $1\%$ (blue) of the mass difference $\delta\Delta M_D ^\text{exp}$. As indicated in the figure, outwards of the contours represents excluded region of the parameter space. Here, the mixing pattern is $V_L^u=V_\text{CKM}^\dagger$ and $V_L^d=\mathbf{1}$. (b) Mass difference as a function of $\cos(\phi+\varphi)$ for $\tan(\phi)=1$ (red), $\tan\phi=50$ (green) and $\tan\phi = 0.01$ (blue). The dashed black line denotes $|\Delta M_D^h|=\delta \Delta M_D^\text{exp}$.}
    \label{fig:higgs_i}
\end{figure}

\begin{itemize}

\item \textbf{Scenario (i):} Consider the scenario in which $V_L^u = V_{\text{CKM}}^\dagger$. For this choice of quark mixing, the contribution of $h$ to the mass differences of $K$, $B_s$, and $B_d$ mesons is exactly zero, so the mass difference $\Delta M_D^h$ becomes the only relevant quantity. On the other hand, the relative uncertainty of the measurement of the $D$-$\bar{D}$ transition is the largest among mesons. This configuration consequently produces the \textit{weakest possible constraints} over the scalar mixing parameters $\phi$ and $\varphi$, establishing it as the most phenomenologically flexible scenario for 3-3-1 model building.

In Fig.~(\ref{fig:higgs_i}a), we present the parameter space dependence of the Higgs-mediated contribution $\Delta M^h_M$ in the $\cos(\phi+\varphi)$--$\tan\phi$ plane. The exclusion regions are demarcated by three contours: red (for $\Delta M_D^h/\delta\Delta M_D^\text{exp} = 1$), green (for $\Delta M_D^h/\delta\Delta M_D^\text{exp} = 0.1$), and blue (for $\Delta M_D^h/\delta\Delta M_D^\text{exp} = 0.01$). Regions interior to the curves are experimentally allowed, while exterior regions are excluded. Notably, for $\Delta M_D^h/\delta\Delta M_D^\text{exp} = 1$ (red curve), no alignment limit constraints arise for $\tan\phi > 20$ or $\tan\phi < 0.05$. Tightened bounds emerge when restricting the Higgs contribution to $10\%$ or $1\%$ of the total experimental uncertainty (green and blue curves, respectively).

In Fig.~(\ref{fig:higgs_i}b), we analyze the $\tan\phi$ dependence across the $\cos(\phi+\varphi)$--$\Delta M^h_M$ parameter space. Three benchmark scenarios are shown: $\tan\phi = 1$ (red), $\tan\phi = 50$ (green), and $\tan\phi = 0.01$ (blue). The distinct trajectories of these curves illustrate how variations in $\tan\phi$ reshape the allowed regions, particularly at extreme values of $\cos(\phi+\varphi)$. This behavior reveals the interplay between quark rotation parametrizations ($V_L^{u,d}$) and scalar alignment conditions in determining viable 3-3-1 model parameter spaces.
\end{itemize}

\begin{figure}[ht!]
    \centering
    \includegraphics[width=0.49\linewidth]{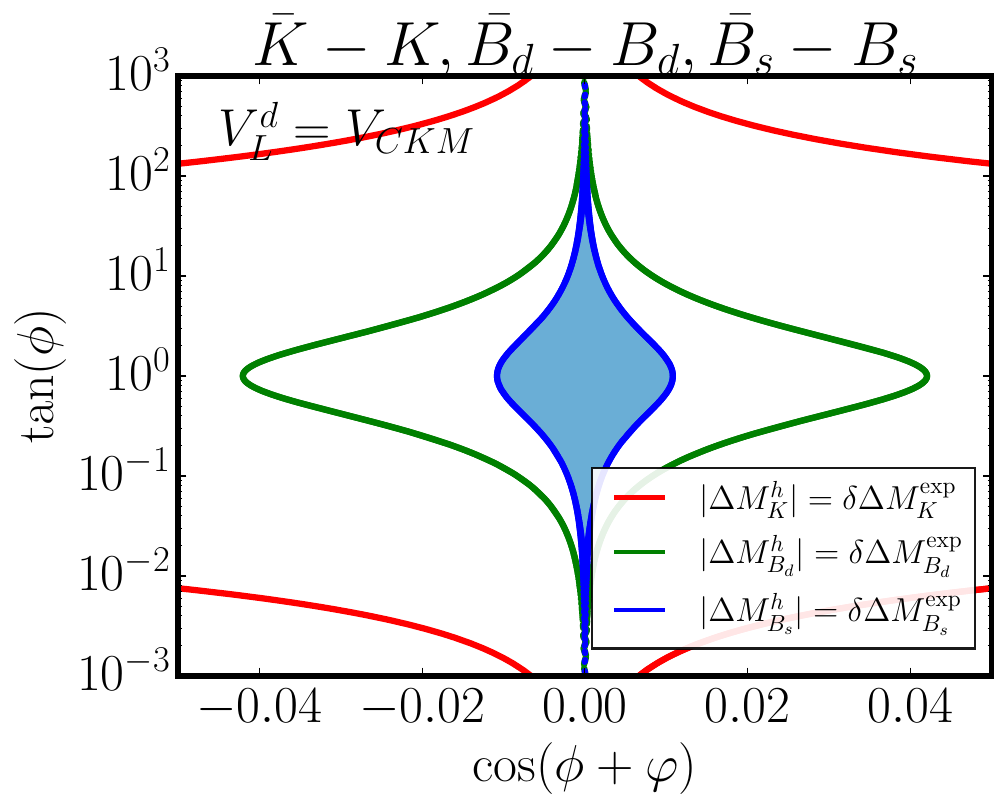}    
    \includegraphics[width=0.49\linewidth]{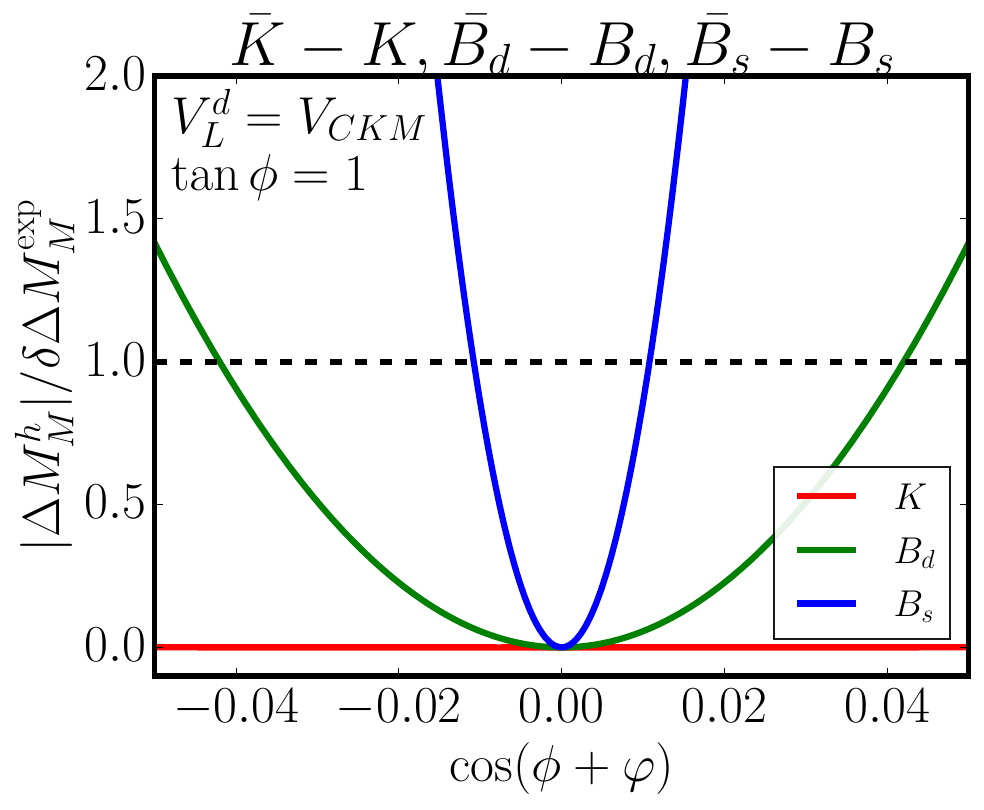}
    \caption{(a) Contours on the $\cos(\phi+\varphi)$--$\tan\phi$ plane fulfilling $100\%$ of the mass differences $\delta\Delta M_M ^\text{exp}$, for mesons $K$ (red), $B_d$ (green) and $B_s$ (blue). As indicated in the figure, outwards of the contours represents excluded region of the parameter space. When combined all meson transition constraints, the only remaining allowed region is indicated by the blue contour. Here, the mixing pattern is $V_L^d=V_\text{CKM}$ and $V_L^u=\mathbf{1}$. (b) Mass difference as a function of $\cos(\phi+\varphi)$ for mesons $K$ (red), $B_d$ (green) and $B_s$ (blue), fixing $\tan \phi=1$. The dashed black line denotes $|\Delta M_M^h|=\delta \Delta M_M^\text{exp}$.}
    \label{fig:higgs_ii}
\end{figure}

\begin{itemize}

\item
\textbf{Scenario (ii):} Consider the scenario in which $V_L^d = V_{\text{CKM}}$. For this choice of quark mixing, the contribution of $h$ to the mass differences of $D$ mesons is exactly zero; then there are FCNC contributions to $K$, $B_s$, and $B_d$ meson transitions. In opposition to Scenario (i), the relative uncertainty of the measurement of the $B_s$-$\bar{B_s}$ transition is the smallest among mesons. This configuration consequently produces another possible constraint over the scalar mixing parameters $\phi$ and $\varphi$, establishing it as a more phenomenologically restrictive scenario for 3-3-1 model building.

In Fig.~(\ref{fig:higgs_ii}a), we analyze the parameter space dependence of the Higgs-mediated contribution $\Delta M^h_M$ in the $\cos(\phi+\varphi)$--$\tan\phi$ plane. Exclusion regions are derived for three meson systems: $K$ (red curve, $\Delta M_K^h/\delta\Delta M_K^\text{exp} = 1$), $B_d$ (green curve, $\Delta M_{B_d}^h/\delta\Delta M_{B_d}^\text{exp} = 1$), and $B_s$ (blue curve, $\Delta M_{B_s}^h/\delta\Delta M_{B_s}^\text{exp} = 1$). The combined constraints from all three systems yield the final allowed region (blue contour), dominated by the stringent $B_s$ transition bounds, which restrict $|\cos(\phi+\varphi)| < 0.01$. In contrast, $K$ and $B_d$ systems permit significantly broader ranges of $\cos(\phi+\varphi)$.

In Fig.~(\ref{fig:higgs_ii}b), we fix $\tan\phi = 1$ to isolate the $\cos(\phi+\varphi)$--$\Delta M^h_M$ parameter space for individual meson systems. Contribution profiles are shown for $K$ (red), $B_d$ (green), and $B_s$ (blue) mesons, with curves representing their respective relative contributions to meson transition experimental bounds.

\end{itemize}

\begin{figure}[ht]
    \centering
    \includegraphics[width=0.49\linewidth]{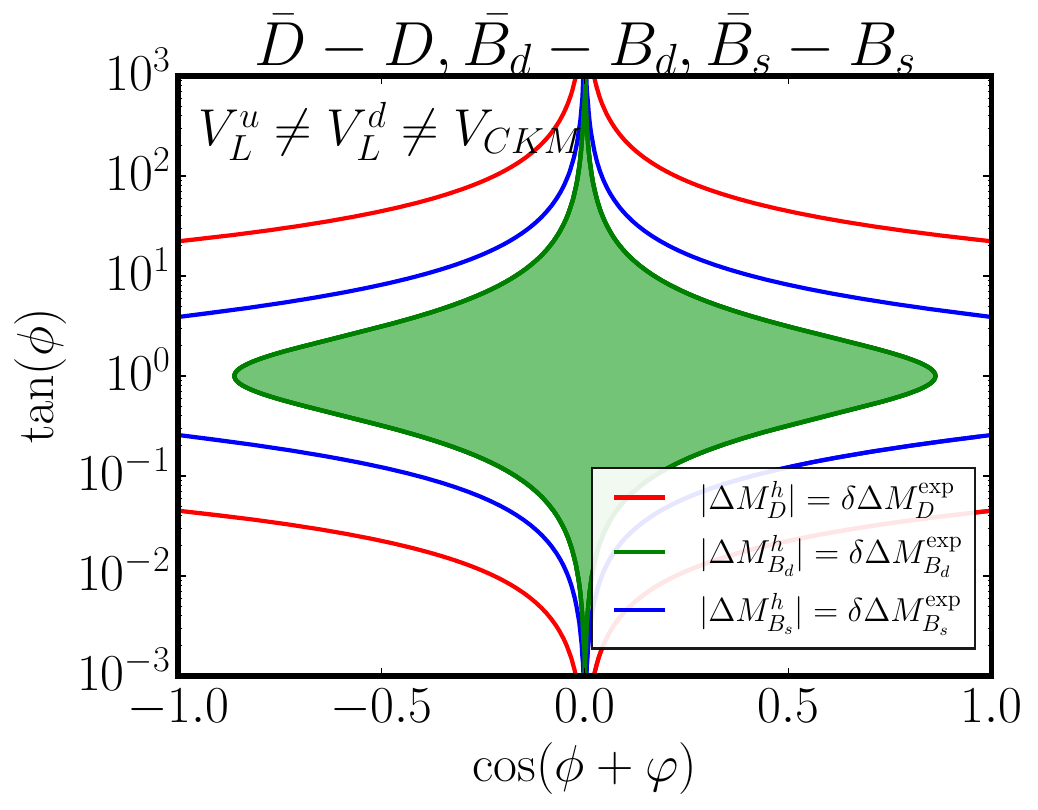}
    \includegraphics[width=0.49\linewidth]{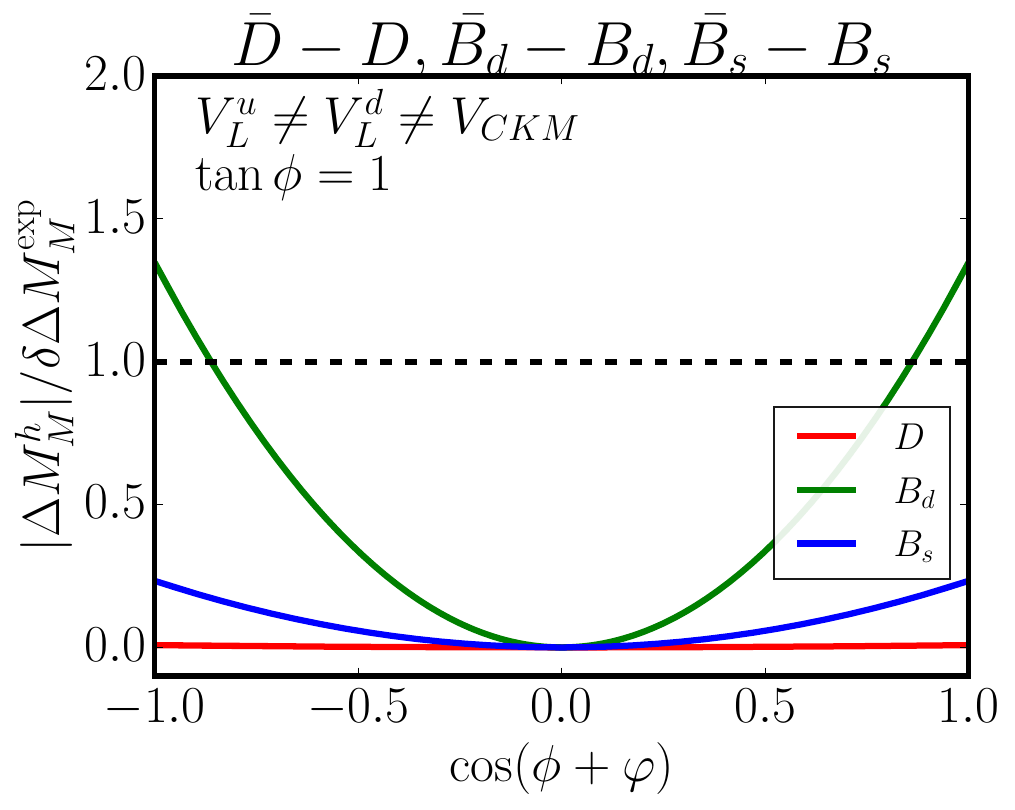}
    \caption{(a) Contours on the $\cos(\phi+\varphi)$--$\tan\phi$ plane fulfilling $100\%$ of the mass differences $\delta\Delta M_M ^\text{exp}$, for mesons $D$ (red), $B_d$ (green) and $B_s$ (blue) ($K$ transition allows all presented plane). As indicated in the figure, outwards of the contours represents excluded region of the parameter space. Here, the mixing pattern $V_L^u$ and $V_L^d$ are explicitly provided in Eq.~\ref{VLDcase3}. (b) Mass difference as a function of $\cos(\phi+\varphi)$ for mesons $D$ (red), $B_d$ (green) and $B_s$ (blue), fixing $\tan \phi=1$. The dashed black line denotes $|\Delta M_M^h|=\delta \Delta M_M^\text{exp}$.}
    \label{fig:higgs_iii}
\end{figure}

\begin{itemize}

\item
\textbf{Scenario (iii):} Now, we turn our attention to an intermediate case, with a less trivial factorization of the CKM matrix. This scenario has non-negligible contributions to three neutral meson--antimeson transitions, $D$, $B_d$ and $B_s$. However, this choice of parametrization suppresses the $K$-$\bar{K}$ transition. We propose the following quark mixing matrices:

\begin{equation}\label{VLUcase3}
    V_L^u =
\scalebox{0.8}{$\displaystyle
\left(
\begin{array}{ccc}
 0.01658\, -\,0.9996\,\mathrm{i} & (-0.1531\,-\,0.7210\, \mathrm{i})\times10^{-4} & (-0.3146\,+\,0.1240\, \mathrm{i})\times 10^{-2} \\
 (-0.9453\,-\,0.2054\, \mathrm{i})\times10^{-4} & -0.7306\,-\,0.6812\,\mathrm{i} & -0.02221\,-\,0.0350 \,\mathrm{i} \\
 (-0.3326\,+\,0.06068\, \mathrm{i})\times 10^{-2} & 0.03055\, +\,0.02801\, \mathrm{i} & -0.5421\,-\,0.8389\, \mathrm{i} \\
\end{array}
\right)$},
\end{equation}
and

\begin{equation}
\label{VLDcase3}
\scalebox{0.8}{$\displaystyle
    V_L^d =\left(
\begin{array}{ccc}
 0.01614\, -\,0.9740\,\mathrm{i} & 0.1644\, +\,0.1532\,\mathrm{i} & (0.1928\, +\,0.4572\,\mathrm{i})\times 10^{-3} \\
 0.003768\, -\,0.2248 \,\mathrm{i} & -0.7126\,-\,0.6643\,\mathrm{i}
 & (0.4807\, +\,0.1342\,\mathrm{i})\times 10^{-3} \\
 -0.1590\times10^{-7}\,-\,0.5663 \,\mathrm{i}\,\times10^{-3} & (0.04655\, -\,0.4242\, \mathrm{i})\times 10^{-3} & -0.5426\,-\,0.8397 \,\mathrm{i} \\
\end{array}
\right)$},
\end{equation}
and this configuration produces intermediate constraints over the scalar mixing parameters $\phi$ and $\varphi$.

In Fig.~(\ref{fig:higgs_iii}a), we analyze the parameter space dependence of the Higgs-mediated contribution $\Delta M^h_M$ in the $\cos(\phi+\varphi)$--$\tan\phi$ plane. Exclusion regions are derived for three meson systems: $D$ (red curve, $\Delta M_D^h/\delta\Delta M_D^\text{exp} = 1$), $B_d$ (green curve, $\Delta M_{B_d}^h/\delta\Delta M_{B_d}^\text{exp} = 1$), and $B_s$ (blue curve, $\Delta M_{B_s}^h/\delta\Delta M_{B_s}^\text{exp} = 1$). The combined constraints from all three systems yield the final allowed region (green contour), dominated by the stringent $B_d$ transition bounds, which restrict $|\cos(\phi+\varphi)| < 0.9$. In contrast, $D$ and $B_s$ systems permit significantly broader ranges of $\cos(\phi+\varphi)$.

In Fig.~(\ref{fig:higgs_iii}b), we fix $\tan\phi = 1$ to isolate the $\cos(\phi+\varphi)$--$\Delta M^h_M$ parameter space for individual meson systems. Contribution profiles are shown for $D$ (red), $B_d$ (green), and $B_s$ (blue) mesons, with curves representing their respective relative contributions to meson transition experimental bounds.
\end{itemize}

The three scenarios analyzed above reveal how the alignment condition $\phi + \varphi = \pi/2$ and the choice of quark mixing parameterization jointly determine the parameter space allowed by the scalar-mediated FCNC. The constraints range from remarkably weak (Scenario i) to extremely stringent (Scenario ii), with intermediate cases (Scenario iii) exhibiting sensitivity to multiple meson systems simultaneously.

Having established the scalar sector constraints, we now examine how these same quark mixing configurations affect the bounds on the $Z^\prime$ gauge boson mass. As we will see, the pattern observed in the scalar analysis---with Scenario (i) yielding the weakest bounds and Scenario (ii) the strongest---persists in the gauge sector, though the relative magnitudes of the constraints differ significantly.

\subsection{\texorpdfstring{$Z^{\prime}$}{Z' } mass estimation using meson transition bounds}
\label{subsec:Zprimebounds}

\begin{figure}[!ht]
    \centering
    \includegraphics[width=0.7\linewidth]{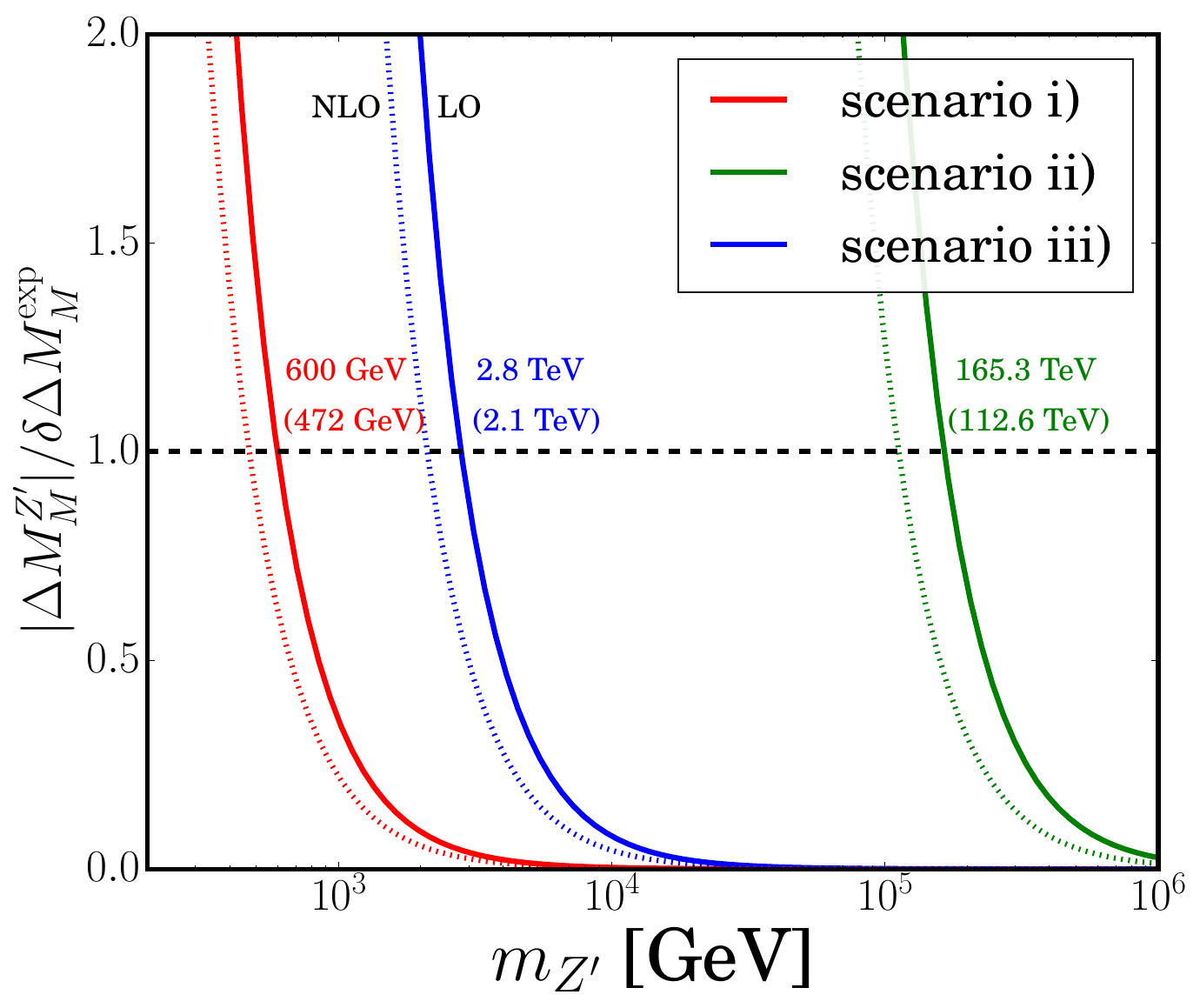}
    \caption{Lower bounds for $m_{Z^\prime}$, presented as the mass difference $\Delta M_M^{Z^\prime}$ as a function of $m_{Z^\prime}$ of the most restrictive meson system of each scenario: $D, B_s$ and $B_d$ for scenarios (i) (red), (ii) (green) and (iii) blue. The full lines represents the LO contributions and the dotted lines represents the NLO contributions.}
    \label{fig:Zprime_bounds}
\end{figure}

After describing the scalar sector contribution to meson transition parameters, it is time for the $Z^\prime$ boson. The  gauge sector of the model  is composed by the standard gauge bosons plus $Z^{\prime}$, two new charged gauge bosons $W^{\prime \pm}$ and two non-hermitian neutral gauge bosons $U^0$ and $U^{0 \dagger}$. For the development of this sector, we refer the reader to Ref.~\cite{Long:1995ctv}. Neglecting mixing among the standard neutral gauge bosons, $Z$ and $Z^{\prime}$, the FCNC processes get mediated by the neutral scalars and $Z^{\prime}$. The  interactions of $Z^{\prime}$ with the quarks that provide FCNC are given by Eqs.~\ref{eq:FCNCZprimeu} and ~\ref{eq:FCNCZprimed}, and the induced  mass differences are given by Eq.~\ref{contribution_Zprime}. We work in the variant I, setting $x=3$, to analyze the three mixing scenarios \textbf{i)}, \textbf{ii)}, and \textbf{iii)} defined in Section~\ref{sec:variants}. Our results, displayed in Fig.~\ref{fig:Zprime_bounds}, map the experimental bounds on the $Z^\prime$ mass ($m_{Z^\prime}$) derived from meson transition data in Tab.~\ref{tab:parameters_table} for leading order (LO) represented as full lines and next-to-leading order (NLO) represented as dotted lines.  

\begin{itemize}
    \item \textbf{Scenario i)} (red curve): The $Z^\prime$ contribution to $D-\bar{D}$ mixing ($\Delta M_D$) yields the weakest constraint. $m_{Z^\prime} \gtrsim 600\;\text{GeV}$ at LO and $m_{Z^\prime} \gtrsim 472\;\text{GeV}$ at NLO, contrasting with conventional expectations~\cite{Buras:2012dp,Cogollo:2012ek,Queiroz:2016gif,Okada:2016whh,CarcamoHernandez:2022fvl, deJesus:2023lvn,Oliveira:2022dav,Buras:2023ldz}. This suppression arises because $V_\text{CKM}^\dagger = V_L^u$ eliminates FCNC in the down-quark sector, leaving only up-type quark processes. Given the comparatively weak experimental limits on $D-\bar{D}$ transitions, this scenario permits a remarkably light $Z^\prime$.  

\item \textbf{Scenario ii)} (green curve): For $B_s-\bar{B}_s$ mixing ($\Delta M_{B_s}$), the $Z^\prime$ mass bound tightens to $m_{Z^\prime} \gtrsim 165.3\;\text{TeV}$ at LO and $112.6$ TeV at NLO, aligning with the literature~\cite{Buras:2012dp,Cogollo:2012ek,Queiroz:2016gif,Okada:2016whh,CarcamoHernandez:2022fvl, deJesus:2023lvn,Oliveira:2022dav,Oliveira:2022dav,Buras:2023ldz}. Here, $V_\text{CKM} = V_L^d$ reintroduces FCNCs in the down-quark sector, subjecting $Z^\prime$ couplings to the experimental constraints on $B_s$ transitions. This explains the dramatic increase in $m_{Z^\prime}$ relative to \textbf{scenario i)}.  

\item \textbf{Scenario iii)} (blue curve): The hybrid mixing pattern affecting $B_d-\bar{B}_d$ transitions ($\Delta M_{B_d}$) yields an intermediate bound of $m_{Z^\prime} \gtrsim 2.8\;\text{TeV}$  at LO and $2.1$ TeV at NLO. This scenario introduces FCNCs across both up- and down-type quarks, with $B_d$ observables dominating the constraints.  

\end{itemize}

Tab.~\ref{tab:zprimemass} summarizes these results. They highlight how CKM parameterization choices in 3-3-1 models critically modulate FCNC visibility, thereby reshaping $Z^\prime$ mass limits. 

\begin{table}[ht]
    \centering
    \caption{Lower bounds on $m_{Z^\prime}$ for each type of meson transition for scenarios \textbf{i)}, \textbf{ii)} and \textbf{iii)}. The limits are presented at LO and NLO outside and inside the parenthesis, respectively.}
    \label{tab:zprimemass}
    \scalebox{0.8}{
    \begin{tabular}{|c|c|c|c|}
        \hline
        \diagbox{\textbf{Quark mixing}}{\textbf{Scenario}} & \textbf{(i)} & \textbf{(ii)} & \textbf{(iii)} \\ 
        \hline
        $V_L^u$ & $V_\text{CKM}^\dagger$ & $\mathbf{1}$ & Eq.~\ref{VLUcase3} \\  
        $V_L^d$ & $\mathbf{1}$ & $V_\text{CKM}$ & Eq.~\ref{VLDcase3} \\
        \hline
        \textbf{Meson} & \textbf{(i)} & \textbf{(ii)} & \textbf{(iii)} \\
        \hline
        $D-\bar{D}$ & $m_{Z^\prime} \geq 600.0(473.0)$ GeV& $-$ & $m_{Z^\prime} \geq 582.3(459.4)$ GeV \\  
        $K-\bar{K}$ & $-$ & $m_{Z^\prime} \geq 6.5(4.8)$ TeV & $m_{Z^\prime} \geq 5.0 (4.9)$ GeV \\  
        $B_d-\bar{B_d}$ & $-$ & $m_{Z^\prime} \geq 49.7(34.7)$ TeV & $m_{Z^\prime} \geq 2.8 (2.1)$ TeV  \\  
        $B_s-\bar{B_s}$ & $-$ & $m_{Z^\prime} \geq 165.3(112.6)$ TeV & $m_{Z^\prime} \geq 1.7(1.3)$ TeV \\  
        \hline
    \end{tabular}
    }
\end{table}

\section{Conclusions and Future Prospects}
\label{sec:conclusion}

FCNC processes are an inherent and unavoidable feature of 3-3-1 models. Three decades of investigation have transformed our understanding of these processes within the 331RHN framework. What began in the late 1990s as isolated studies of $Z^\prime$-mediated FCNC has evolved into a comprehensive theoretical picture that recognizes the indispensable interplay between gauge and scalar sectors. Throughout this review, we have examined in detail the intricate relationship between gauge and scalar contributions to FCNC interactions. The most significant advances from recent literature can be summarized as follows:

\paragraph{Role of the SM Higgs.}
The most significant conceptual advance recently was the recognition that the SM-like Higgs boson $h$ plays a decisive role in FCNC processes. Unlike $Z^\prime$ contributions, which can be suppressed by increasing the gauge boson mass, scalar-mediated FCNC scale as $1/m_h^2$ and are intrinsically related to the quark mass spectrum and CKM structure. This has deep implications: any parametrization of the quark mixing matrices $V_L^{u,d}$ is phenomenologically viable only if the SM-like Higgs contribution to meson transitions remains within experimental bounds. Studies that neglect this contribution---as was standard practice until recently---may lead to erroneous conclusions regarding the allowed parameter space.

\paragraph{The Alignment Limit.}
The identification of the alignment condition $\phi + \varphi = \pi/2$ represents another important moment in 331RHN phenomenology. Within this limit, the SM-like Higgs decouples from FCNC at tree level, naturally suppressing the most stringent constraints. This discovery reveals two contrasting phenomenological scenarios:
\begin{itemize}
\item \textbf{Up-quark sector FCNC} (Scenario i, $V_L^u = V_{\text{CKM}}^\dagger$, $V_L^d = \mathbf{1}$): FCNC processes are confined to $D$--$\bar{D}$ mixing, yielding weak bounds that permit $m_{Z^\prime} \gtrsim 600$ GeV at leading order ($\sim 470$ GeV at NLO). This scenario represents \textit{the most phenomenologically flexible} configuration.

\item \textbf{Down-quark sector FCNC} (Scenario ii, $V_L^d = V_{\text{CKM}}$, $V_L^u = \mathbf{1}$): FCNC arise in $K$, $B_d$, and $B_s$ systems, with $B_s$ mixing imposing the most stringent constraint of $m_{Z^\prime} \gtrsim 165$ TeV at LO ($\sim 113$ TeV at NLO). This pushes the 331 scale far beyond direct LHC reach, making the model testable only through precision measurements and indirect probes.

\item \textbf{Hybrid scenarios} (Scenario iii): Intermediate configurations where both sectors contribute yield bounds $\sim$ $ 470$ GeV--$165$ TeV, depending on the specific mixing structure. 
\end{itemize}

The two-orders-of-magnitude variation in $Z^\prime$ mass bounds between these scenarios---from hundreds of GeV to hundreds of TeV---marks a fundamental feature of 331RHN phenomenology: \textit{the viable parameter space is highly sensitive to the assumed quark mixing structure and scalar alignment configuration}.

\paragraph{Family Discrimination.}
The question of which quark family transforms as a triplet under SU$(3)_L$ has been addressed through meson transition observables. Studies of pseudoscalar contributions to $K$--$\bar{K}$ mixing~\cite{Oliveira:2022vjo} demonstrate that variant I (third family as triplet) yields significantly weaker bounds ($m_A \gtrsim 2.6$ TeV) compared to variants II and III (first or second family as triplet, requiring $m_A \gtrsim 17$--$20$ TeV). Combined with the phenomenological observation that the top quark's rapid decay prevents it from participating in meson transitions, these results provide strong---though not yet definitive---evidence favoring the third family assignment.

\paragraph{$Z$--$Z^\prime$ Mixing Effects.}
Although the mixing angle $\theta_{331}$ is suppressed by $(v_{\text{SM}}/v_{\chi^\prime})^2$, the induced flavor-changing couplings of the SM-like $Z_1$ boson provide complementary constraints. Analyses incorporating contributions from $h$, $Z^\prime$, and $Z_1$ to neutral meson transitions yield bounds $\theta_{331} \lesssim 6 \times 10^{-3}$~\cite{Oliveira:2022dav}, consistent with previous estimates and compatible with top quark decay constraints from ATLAS~\cite{ATLAS:2017tas}.

\subsection{Open Questions and Testability}

Despite substantial progress, there are fundamental questions that remain unresolved:

\paragraph{Validity of the Alignment Limit.}
While the alignment condition $\phi + \varphi = \pi/2$ provides a mechanism for suppressing Higgs-mediated FCNC, its physical realization requires specific relationships between scalar potential parameters. Systematic studies are needed to determine whether this configuration is naturally achieved through radiative corrections or fine-tuning.  Precision measurements of Higgs couplings at future colliders could provide indirect constraints on alignment through modified flavor-diagonal couplings.

\paragraph{Right-Handed Mixing Matrices.}
Throughout this review, we have worked in a basis where right-handed quarks are diagonal, $V_R^{u,d} = \mathbf{1}$. However, the most general case allows for non-trivial $V_R^{u,d}$ matrices that introduce additional sources of FCNC through modified scalar couplings. The phenomenological impact of activating right-handed mixing remains largely unexplored.

The phenomenological viability of 331RHN models depends critically on the $Z^\prime$ mass scale, which varies from hundreds of GeV to hundreds of TeV depending on the quark mixing scenario and alignment configuration:

\paragraph{Low-Scale Scenarios ($m_{Z^\prime} \sim $ few hundred GeV).}
If alignment is realized with up-quark sector FCNC, direct $Z^\prime$ production via Drell-Yan processes at HL-LHC.

\paragraph{Intermediate-Scale Scenarios ($m_{Z^\prime} \sim $ few TeV).}
HL-LHC can probe $Z^\prime$ masses up to $6$--$7$ TeV in dilepton channels, while precision flavor measurements at Belle II and LHCb, combined with Higgs coupling deviations at future lepton colliders, offer complementary constraints.

\paragraph{High-Scale Scenarios ($m_{Z^\prime} \gtrsim 100$ TeV).}
Direct discovery becomes very difficult in current and near-future experiments. However, ultra-precision kaon experiments such as NA62 and KOTO~\cite{Romano:2024dcw}, which aim to measure  $K \to \pi \nu \bar{\nu}$ at $\mathcal{O}(10^{-13})$ sensitivity, together with global fits using effective field theory approaches can indirectly constrain or reveal the 3-3-1 scale.

\subsection{Final Remarks}

The 331RHN model stands as one of the most compelling extensions of the SM, offering explanations for the number of fermion families and electric charge quantization within a single coherent framework. Its rich flavor phenomenology---with tree-level FCNC unavoidable by construction---makes it simultaneously highly predictive and tightly constrained.

The central lesson from three decades of FCNC studies is clear: \textbf{the scalar sector cannot be ignored}. Any comprehensive phenomenological analysis must account for contributions from the SM-like Higgs $h$, the heavier scalars $H$ and $A$, and the $Z^\prime$ gauge boson on equal footing. The discovery of the alignment limit provides a theoretically motivated framework for organizing this complexity, but many questions remain about its physical origin and the broader structure of the quark mixing matrices.

Looking forward, the interplay between direct searches at colliders and precision measurements in flavor physics will be crucial for testing 331RHN models. If nature has chosen the low-scale scenario with alignment, discovery could be imminent at the LHC or future colliders. If the high-scale scenario is realized, indirect probes through ultra-precise flavor observables and effective field theory techniques offer the most promising path forward. In either case, the next decade of experimental and theoretical developments promises to significantly clarify---and possibly confirm---the 331RHN framework as the correct description of physics beyond the SM.

The journey from simple $Z^\prime$-mediated constraints in the 1990s to today's sophisticated understanding of scalar alignment and multi-mediator FCNC illustrates the power of sustained theoretical investigation informed by experimental progress. As we enter an era of precision measurements at Belle II, LHCb upgrades, and future colliders, the 331RHN model---and particularly its Variant I with right-handed neutrinos---remains a prime candidate for explaining the mysteries of flavor physics while addressing fundamental questions left unanswered by the SM.

\section*{Acknowledgments}
J.P.P. is supported by the National Natural Science Foundation of China (12425506, 12375101, 12090060, 12090064) and the SJTU Double First Class start-up fund (WF220442604).
V.O. is supported by CIDMA under the Portuguese Foundation for Science and Technology (FCT) Multi-Annual Financing Program for R\&D Units (grants UID/4106/2025, UID/PRR/4106/2025) and by FCT doctoral grant PRT/BD/154629/2022. V.O. thanks Lund University for hospitality. P.E. and C.A.S.P. are supported by CNPq grants 151612/2024-2 and 311936/2021-0, respectively. P.E. thanks IIP at UFRN for its hospitality during the final stages of this work.

\appendix

\section{Coefficients of effective lagrangians}
\label{app:coefficients}

The coefficients $C_{M,h}^{R,L},\,\,C_{M,H}^{R,L}$ and $C_{M,A}^{R,L}$ presented in Eqs. \ref{contribution_h}, \ref{contribution_A}, and \ref{contribution_H} are given by the following expressions:
\begin{eqnarray}
    C_{D,h}^R &=& \frac{m_c}{v}  \left(V_L^u\right)_{x2}^* \left(V_L^u\right)_{1x}, \nonumber \\
    C_{D,h}^L &=& \frac{m_u}{v}   \left(V_L^u\right)^*_{1x} \left(V_L^u\right)_{x2}, \nonumber \\
    C_{K,h}^R &=& \frac{m_d}{v} \left(V_L^d\right)^*_{x1} \left(V_L^d\right)_{2x}, \nonumber \\
    C_{K,h}^L &=&  \frac{m_s}{v}  \left(V_L^d\right)^*_{2x} \left(V_L^d\right)_{x1}, \nonumber \\
    C_{B_d,h}^R &=& \frac{m_d}{v} \left(V_L^d\right)_{x1}^* \left(V_L^d\right)_{3x}, \nonumber \\
    C_{B_d,h}^L &=&  \frac{m_b}{v}  \left(V_L^d\right)^*_{3x} \left(V_L^d\right)_{x1}, \nonumber \\
    C_{B_s,h}^R &=& \frac{m_s}{v} \left(V_L^d\right)_{x2}^* \left(V_L^d\right)_{3x}, \nonumber \\
    C_{B_s,h}^L &=&  \frac{m_b}{v}  \left(V_L^d\right)^*_{3x} \left(V_L^d\right)_{x2},
\end{eqnarray}
\begin{eqnarray}
    C_{D,H}^R &=& \frac{m_c}{v} \sin (\phi-\varphi)\left(V_L^u\right)_{x2}^* \left(V_L^u\right)_{1x}, \nonumber \\
    C_{D,H}^L &=& \frac{m_u}{v} \sin (\phi-\varphi)\left(V_L^u\right)^*_{1x} \left(V_L^u\right)_{x2}, \nonumber \\
    C_{K,H}^R &=& \frac{m_d}{v} \sin (\phi+\varphi) \left(V_L^d\right)^*_{x1} \left(V_L^d\right)_{2x}, \nonumber \\
    C_{K,H}^L &=&  \frac{m_s}{v} \sin (\phi+\varphi) \left(V_L^d\right)^*_{2x} \left(V_L^d\right)_{x1}, \nonumber \\
    C_{B_d,H}^R &=& \frac{m_d}{v} \sin (\phi+\varphi) \left(V_L^d\right)_{x1}^* \left(V_L^d\right)_{3x}, \nonumber \\
    C_{B_d,H}^L &=&  \frac{m_b}{v} \sin (\phi+\varphi) \left(V_L^d\right)^*_{3x} \left(V_L^d\right)_{x1}, \nonumber \\
    C_{B_s,H}^R &=& \frac{m_s}{v} \sin (\phi+\varphi) \left(V_L^d\right)_{x2}^* \left(V_L^d\right)_{3x}, \nonumber \\
    C_{B_s,H}^L &=&  \frac{m_b}{v} \sin (\phi+\varphi)  \left(V_L^d\right)^*_{3x} \left(V_L^d\right)_{x2},
\end{eqnarray}
\begin{eqnarray}
    C_{D,A}^R &=& \frac{m_c}{v} (\tan\phi+\cot\phi) \left(V_L^u\right)_{x2}^* \left(V_L^u\right)_{1x}, \nonumber \\
    C_{D,A}^L &=& \frac{m_u}{v} (\tan\phi+\cot\phi)  \left(V_L^u\right)^*_{1x} \left(V_L^u\right)_{x2}, \nonumber\\
    C_{K,A}^R &=& \frac{m_d}{v} (\tan\phi-\cot\phi) \left(V_L^d\right)^*_{x1} \left(V_L^d\right)_{2x}, \nonumber \\
    C_{K,A}^L &=&  \frac{m_s}{v} (\tan\phi-\cot\phi) \left(V_L^d\right)^*_{2x} \left(V_L^d\right)_{x1}, \nonumber \\
    C_{B_d,A}^R &=& \frac{m_d}{v} (\tan\phi-\cot\phi) \left(V_L^d\right)_{x1}^* \left(V_L^d\right)_{3x}, \nonumber \\
    C_{B_d,A}^L &=&  \frac{m_b}{v} (\tan\phi-\cot\phi) \left(V_L^d\right)^*_{3x} \left(V_L^d\right)_{x1}, \nonumber \\
    C_{B_s,A}^R &=& \frac{m_s}{v}(\tan\phi-\cot\phi) \left(V_L^d\right)_{x2}^* \left(V_L^d\right)_{3x}, \nonumber  \\
    C_{B_s,A}^L &=&  \frac{m_b}{v} (\tan\phi-\cot\phi) \left(V_L^d\right)^*_{3x} \left(V_L^d\right)_{x2},
\end{eqnarray}
where $x$ takes values depending on the variant of the 331RHN model as described in Tab.~\ref{tab:x_parameters}.

\section{Renormalization group evolution of the Wilson coefficient at one-loop}
\label{app:RGcorr}

In this Appendix we present RG evolution of the $\Delta F = 2$ Wilson coefficients from the $Z^\prime$ mass scale to the relevant meson scales.

From Eq.~\ref{eq:FCNCZprimed}, the effective Lagrangian after integrating out the $Z^\prime$ is given by
\begin{equation}
\mathcal{L}_{\text{eff}}^{\Delta F=2} = \frac{g^{\prime 2}}{8m_{Z^\prime}^2} \sum_{i,j} (V_L^{d})_{3i}^* (V_L^{d})_{3j} (\bar{d}_i \gamma^\mu P_L d_j)^2.
\end{equation}
This gives the Wilson coefficient at the matching scale $\mu = m_{Z^\prime}$:
\begin{equation}
C_1(m_{Z^\prime}) = \frac{g^{\prime 2}}{8m_{Z^{\prime}}^2} (V_L^{d})_{3i}^* (V_L^{d})_{3j}.
\end{equation}
The expressions for up-like quarks is analogous.

The RG evolution of the Wilson coefficient is governed by:
\begin{equation}
\mu \frac{d}{d\mu} C_1(\mu) = \gamma_1(\alpha_s) C_1(\mu).
\end{equation}

The solution at leading-log order is:
\begin{equation}
C_1(\mu_L) = \left[\frac{\alpha_s(\mu_H)}{\alpha_s(\mu_L)}\right]^{\gamma_1^{(0)}/(2\beta_0)} C_1(\mu_H),
\end{equation}
where $\mu_H = m_{Z^\prime}$ and $\mu_L$ is the low-energy scale.

For the operator $\mathcal{O}_1 = (\bar{d}_i \gamma^\mu P_L d_j)^2$:
\begin{equation}
\gamma_1^{(0)} = 6 C_F = 6 \times \frac{4}{3} = 8.
\end{equation}

The QCD beta function coefficient is:
\begin{equation}
\beta_0 = \frac{11N_c - 2n_f}{3} = \frac{33 - 2n_f}{3}.
\end{equation}

For $m_{Z^\prime} > m_t$, we must include threshold matching (supposing there are only the standard quarks):
\begin{itemize}
\item For $\mu > m_t$: $n_f = 6$, $\beta_0 = 7$
\item For $m_b < \mu < m_t$: $n_f = 5$, $\beta_0 = 23/3$
\item For $m_c < \mu < m_b$: $n_f = 4$, $\beta_0 = 25/3$
\item For $\mu < m_c$: $n_f = 3$, $\beta_0 = 9$
\end{itemize}

The complete RG factor for $B_{d,s}$ mixing ($\mu_L \approx m_b$) is:

For $m_{Z^\prime} \gg m_t$:
\begin{equation}
r_{RG} = \left[\frac{\alpha_s(m_t)}{\alpha_s(m_b)}\right]^{12/23} \left[\frac{\alpha_s(m_{Z^\prime})}{\alpha_s(m_t)}\right]^{4/7}
\end{equation}

The one-loop RG solution for $\alpha_s$ is:
\begin{equation}
\alpha_s(\mu_2) = \frac{\alpha_s(\mu_1)}{1 + \frac{\beta_0 \alpha_s(\mu_1)}{2\pi} \ln\left(\frac{\mu_2}{\mu_1}\right)}
\end{equation}

Using $\alpha_s(m_Z) = 0.118$, $\alpha_s(m_b) = 0.22$, $\alpha_s(m_t) = 0.108$, we have that:
\begin{align}
&m_{Z^\prime} = 1 \text{ TeV}:  \quad \sqrt{r_{RG}} \approx 0.80, \\
&m_{Z^\prime} = 10 \text{ TeV}:  \quad \sqrt{r_{RG}} \approx 0.73 ,\\
&m_{Z^\prime} = 100 \text{ TeV}:  \quad \sqrt{r_{RG}} \approx 0.69.
\end{align}

Including RG effects, the expressions for the mass differences become:
\begin{equation}
\Delta M_{M}^\text{new}= r_{RG} \times \Delta M_{M}^\text{old}.
\end{equation}

Then, the NLO bounds are:
\begin{equation}
m_{Z^{\prime}}=\sqrt{\frac{ r_{RG} \times \frac{4\sqrt{2}G_F c_W^4}{3-4s_W^2} \left|  \left(V_L^{u,d}\right)^*_{b3}\left(V_L^{u,d}\right)_{3a} \right|^2 m_Z^2 f_{M}^2 B_{M} m_{M}}{ \Delta M_{12}^\text{exp}}}.
\end{equation}

This relaxes the bounds on $m_{Z^\prime}$ by:
\begin{equation}
m_{Z^\prime}^{\text{with RG}} =\sqrt{r_{RG}} m_{Z^\prime}^{\text{without RG}}.
\end{equation}

For example: 
\begin{align}
m_{Z^\prime} = 1 \text{ TeV} &\to  0.8 \text{ TeV}, \\
m_{Z^\prime} = 10 \text{ TeV} &\to  7.3 \text{ TeV}, \\
m_{Z^\prime} = 100 \text{ TeV} &\to  68.8 \text{ TeV}.
\end{align}

\bibliography{references}

\end{document}